\documentclass[aps,prd,12pt,nofootinbib,notitlepage]{revtex4-1}

\usepackage{amsfonts}
\usepackage{amsmath,epsfig,bm}
\usepackage{graphicx}
\usepackage{bm}
\usepackage{color}
\renewcommand{\vec}[1]{\mbox{\boldmath$#1$}}

\usepackage{tabularx} 

\usepackage{times}

\newcommand{\be}{\begin{equation}}
\newcommand{\ee}{\end{equation}}
\newcommand{\ba}{\begin{eqnarray}}
\newcommand{\ea}{\end{eqnarray}}
\newcommand{\bd}{\begin{displaymath}}
\newcommand{\ed}{\end{displaymath}}

\def\thalf{{\textstyle{\frac{1}{2}}}}
\def\tthalf{{\textstyle{\frac{3}{2}}}}

\def\oneqt{{\textstyle{\frac{1}{4}}}}
\def\ttqt{{\textstyle{\frac{3}{4}}}}

\def\ones{{\textstyle{\frac{1}{6}}}}

\def\rt3{\sqrt{3}}
\def\rt6{\sqrt{6}}
\def\mL2{(m_{\phi}L)^2}

\begin{document}

\title{{\bf Relaxation Time for Strange Quark Spin in Rotating Quark-Gluon Plasma}}
\author{Joseph I. Kapusta$^1$, Ermal Rrapaj$^{1,2}$, and Serge Rudaz$^1$}
\affiliation{$^1$School of Physics and Astronomy, University of Minnesota, Minneapolis, Minnesota 55455, USA \\
$^2$Department of Physics, University of California, Berkeley, CA 94720, USA}

\vspace{.3cm}

\parindent=20pt

\begin{abstract}
Experiments at the Relativistic Heavy Ion Collider (RHIC) have measured the net polarization of $\Lambda$ and $\bar{\Lambda}$ hyperons and attributed it to a coupling between their spin and the vorticity of the fluid created in heavy ion collisions.  Equipartition of energy is generally assumed, but the dynamical mechanism which polarizes them has yet to be determined.  We consider two such mechanisms: vorticity fluctuations and helicity flip in scatterings between strange quarks and light quarks and gluons.  With reasonable parameters both mechanisms lead to equilibration times orders of magnitude too large to be relevant to heavy ion collisions.  Our conclusion is that strange quark spin or helicity is unchanged from the time they are created to the time they hadronize.  A corollary is that vorticity fluctuations do not affect the hyperon spin either.
\end{abstract}
\date{\today}

\maketitle

\section{Introduction}
\label{SecI}

Experiments at the Relativistic Heavy Ion Collider (RHIC) and at the Large Hadron Collider (LHC) have provided an abundance of data on the hot, dense matter created in heavy ion collisions \cite{QMseries}.  Among these data are the coefficients of a Fourier expansion in the azimuthal angle for a variety of physical observables.  The data provide strong evidence for collective expansion of the hot, dense matter and provide information on transport coefficients such as the shear viscosity \cite{whitepaper}.  In addition, the polarization of $\Lambda$ and $\bar{\Lambda}$ hyperons was proposed as yet another observable that provides information on collective flow, in particular vorticity \cite{Wang1,Becattini1}.  The vorticity arises in non-central heavy ion collisions where the produced matter has considerable angular momentum.  The spins of the $\Lambda$ and $\bar{\Lambda}$ couple to the vorticity, resulting in a splitting in energy between particles with spin parallel and antiparallel to the vorticity.  The decay products of these hyperons are used to infer their polarizations.  Measurements of the polarizations have been made by the STAR collaboration from the lowest to the highest beam energies at RHIC \cite{FirstSTAR,Nature,SecondSTAR}, noting that RHIC produces matter with the highest vorticity ever observed.  

The standard picture of $\Lambda$ and $\bar{\Lambda}$ polarization in non-central heavy ion collisions assumes equipartition of energy \cite{Becattini2,Becattini3}.  The spin-vorticity coupling is the same for baryons and antibaryons, which is approximately what is observed.  The relatively small difference was studied in Ref. \cite{CKW} and will not be addressed here.   Similar to the difficult question of how the quarks and gluons come to thermal equilibrium is the dynamical mechanism by which the hyperons become polarized.  Within the quark model the spin of the $\Lambda$ is carried by the strange quark \cite{Jennings1,Cohen}.  One possibility is that the $s$ and $\bar s$ quarks become polarized in the quark-gluon plasma phase and pass that poalrization on to the $\Lambda$ and $\bar{\Lambda}$ during hadronization.  We shall estimate the relaxation rates and times for the strange quark spin to come to equilibrium with the vorticity in an idealized situation of a rotating quark-gluon plasma. 

Suppose that the strange quark spin is not in equilibrium.  We consider two mechanisms by which it would be brought back to equilibrium.  The first mechanism recognizes that there will be fluctuations in the direction and magnitude of the vorticity in heavy ion collisions.  These fluctuations will drive the spins back towards equilibium, just as fluctuations around a constant magnetic field drive electron spins towards equilibrium.   The second mechanism considers the scattering of massive strange quarks with massless up and down quarks and gluons in the plasma.  Since helicity is conserved in QCD interactions when the quark is massless, helicity flip can only occur when the quark has a mass.  At the scales of interest, the current quark masses of the up and down quarks are in the range from 4 to 7 MeV and may be considered massless.  If their helicities are out of equilibrium they cannot be brought into equilibrium by scattering.  They evolve without change.  The current quark mass of the strange quark is in the range of 100 to 120 MeV.  Their equilibration times will be nonzero due to scattering.

The outline of the paper is as follows.  In Sec. \ref{SecII} we review the use of tetrads for spin-1/2 fermions in an accelerated system, in particular for a rotating system.  In Sec. \ref{SecIII} we find the eigenvalues and eigenvectors for spin-1/2 fermions in a model Hamiltonian where the orbital angular momentum is small enough that it may be neglected.  In Sec. \ref{SecIV} we review fluctuations at the level of second order response theory as linear response theory is insufficient.  We apply this to a massive Pauli particle in Sec. \ref{SecV}, to a massless Dirac particle in Sec. \ref{SecVI}, and to a massive Dirac particle in Sec. \ref{SecVII}.  In the latter two sections we restrict our attention to the case when the momentum is parallel or anti-parallel to the direction of vorticity, both for reasons of simplicity and because we neglect orbitatal angular momentum in these sections.  In Sec. \ref{SecVIII} we apply kinetic theory to the rate of helicity flip of strange quarks and anti-quarks using lowest order QCD perturbation theory.  In Sec. \ref{SecIX} we present numerical results, and in Sec. \ref{SecX} we give our conclusions.  Some useful commutation and anti-commutation relations are recalled in the appendix.  The appendix also contains many of the tedious mathematical details.

Readers mainly interested in the results and not the mathematical details may wish to read Secs. \ref{SecIX} and \ref{SecX} first.

\section{Tetrads and Spin}
\label{SecII}

Consider a fluid element undergoing linear acceleration, expansion, and rotation.  Although we are mostly interested in rotation the tetrad formalism is able to handle all types.  See Ref. \cite{Yepez} for a clear review of the topic.\footnote{Equation (188) in this reference is consistent with the literature.  The more compact expression given in Eq. (194d) is wrong since it gives a different result for the $\Gamma_{\mu}$ derived below.}  The idea is to set up an inertial coordinate system at rest with respect to a fluid element at every space-time point.  Let $x^{\mu}$ represent the space-time coordinates of an observer at rest in the fluid element and $\xi^a$ the coordinates of an inertial frame.  Then
\be
 g_{\mu\nu}(x) dx^{\mu} dx^{\nu} = \eta_{ab} d\xi^a d\xi^b  \,.
\ee
When there is no cause for confusion we use Greek indices for the $x$-coordinates, Latin indices $a, b, ...$ for the $\xi$-coordinates, and Latin indices $i, j, ...$ for spatial indices.  The Minkowski metric is $\eta_{ab} = {\rm diag}(1,-1,-1,-1)$.  The tetrad is defined as
\be
e_{\mu}^{\;\;a}(x) = \frac{\partial \xi^a}{\partial x^{\mu}}
\ee
while the inverse tetrad is 
\be
e^{\mu}_{\;\;a}(x) = g^{\mu\nu}(x) \eta_{ab} e_{\nu}^{\;\;b}(x) \,.
\ee
Note that Greek indices are raised and lowered with $g_{\mu\nu}(x)$ and its inverse, while Latin indices are raised and lowered by $\eta_{ab}$ and its inverse.  The tetrads obey the orthogonality properties
\ba
e_{\mu}^{\;\;a}(x) e^{\mu}_{\;\;b}(x) &=& \delta^a_b \nonumber \\
e^{\mu}_{\;\;a}(x) e_{\nu}^{\;\;a}(x) &=& \delta^{\mu}_{\nu} \,.
\ea

In analogy to the affine connection $\Gamma^{\lambda}_{\mu\nu}$ there is a spin connection $\omega_{\mu \;\;b}^{\;\;a}$ which is used to take covariant derivatives of spinors.  It satisfies the equation
\be
\omega_{\mu \;\;b}^{\;\;a} = e_{\nu}^{\;\;a} e^{\lambda}_{\;\;b} \Gamma^{\nu}_{\mu\lambda} - e^{\lambda}_{\;\;b} \partial_{\mu} e_{\lambda}^{\;\;a} \,.
\ee
The Dirac matrices $\hat{\gamma}^{\mu}(x)$ become space-time dependent.  They are obtained from the usual Dirac matrices $\gamma^a$ by
\be
\hat{\gamma}^{\mu}(x) = e^{\mu}_{\;\;a}(x) \gamma^a \,.
\ee
They satisfy
\be
\hat{\gamma}^{\mu} \hat{\gamma}^{\nu} + \hat{\gamma}^{\nu} \hat{\gamma}^{\mu} = 2 g^{\mu\nu}
\ee
compared to
\be
\gamma^{a} \gamma^{b} + \gamma^{b} \gamma^{a} = 2 \eta^{ab} \,.
\ee
One finds that the gradient of a spinor is replaced by a covariant derivative.
\be
\partial_{\mu} \psi \rightarrow D_{\mu} \psi = \left(\partial_{\mu} + \Gamma_{\mu} - i e A_{\mu}\right) \psi
\ee
Here an electromagnetic vector potential is included for reference.  The symbol $\Gamma_{\mu}$ is also called the spin connection, which is confusing in several ways.  The Dirac equation is \cite{Brill}
\be
i \hat{\gamma}^{\mu}(x) D_{\mu} \psi - m \psi = 0 \,.
\ee
The spin connection is
\be
\Gamma_{\mu} = -\thalf \omega_{\mu a b} S^{ab}
\label{Yepezspinconnect}
\ee
where 
\be
S^{ab} = \frac{i}{2} \sigma^{ab} \;\;\; {\rm and} \;\;\; \sigma^{ab} = \frac{i}{2} [\gamma^a,\gamma^b] \,.
\ee

Consider a region of space where a fluid element is rotating in an anti-clockwise sense around the $z$ axis with angular speed $\omega$ which may be considered constant within that region.  Here we follow Ref. \cite{Hehl} and choose the tetrad as the $4\times 4$ matrix
\be
e_{\mu}^{\;\;a}(x) =
\begin{pmatrix}
1 & v_x & v_y & 0 \\
0 & 1 & 0 & 0 \\
0 & 0 & 1 & 0 \\
0 & 0 & 0 & 1 \\
\end{pmatrix}
\ee
where $v_x \equiv -\omega y$ and $v_y \equiv \omega x$.  From this is it straightforward to find the metric
\be
g_{\mu\nu}(x) =
\begin{pmatrix}
1 -v^2 & -v_x & -v_y & 0 \\
-v_x & -1 & 0 & 0 \\
-v_y & 0 & -1 & 0 \\
0 & 0 & 0 & -1 \\
\end{pmatrix} \, ,
\ee
the inverse metric
\be
g^{\mu\nu}(x) =
\begin{pmatrix}
1 & -v_x & -v_y & 0 \\
-v_x & -1+v_x^2 & v_x v_y & 0 \\
-v_y & v_x v_y & -1+v_y^2 & 0 \\
0 & 0 & 0 & -1 \\
\end{pmatrix} \, ,
\ee
and the inverse tetrad
\be
e^{\mu}_{\;\;a}(x) =
\begin{pmatrix}
1 & 0 & 0 & 0 \\
-v_x & 1 & 0 & 0 \\
-v_y & 0 & 1 & 0 \\
0 & 0 & 0 & 1 \\
\end{pmatrix} \, .
\ee

To get the spin connection it is useful to have
\be
e_{\nu a}(x) =
\begin{pmatrix}
1 & -v_x & -v_y & 0 \\
0 & -1 & 0 & 0 \\
0 & 0 & -1 & 0 \\
0 & 0 & 0 & -1 \\
\end{pmatrix} \,.
\ee
The nonzero components of the affine connection are
\ba
\Gamma^1_{00} &=& \omega v_y \nonumber \\
\Gamma^2_{00} &=& -\omega v_x \nonumber \\
\Gamma^2_{01} &=& \omega \nonumber \\
\Gamma^1_{02} &=& -\omega \,.
\ea
The nonzero components of $\omega_{\mu a b}$ are
\be
\omega_{012} = -\omega_{021} = \omega \,.
\ee
Hence the only nonzero component of $\Gamma_{\mu}$ is
\be
\Gamma_0 = -\frac{i}{2} \omega \Sigma_3
\ee
where 
\be
\Sigma_j = \begin{pmatrix}
\sigma_j & 0 \\
0 & \sigma_j \\
\end{pmatrix} \,.
\ee 
Finally the Dirac matrices are
\ba
\hat{\gamma}^0 &=& \gamma^0 \nonumber \\
\hat{\gamma}^1 &=& \gamma^1 - v_x \gamma^0 \nonumber \\
\hat{\gamma}^2 &=& \gamma^2 - v_y \gamma^0 \nonumber \\
\hat{\gamma}^3 &=& \gamma^3
\ea

The single particle Hamiltonian can be found by writing the Dirac equation in the form $i \partial_0 \psi = H \psi$ with the result
\be
H = \beta m -i \alpha^j \partial_j + i \omega (x \partial_2 - y \partial_1) - \thalf \omega \Sigma_3
\label{spH}
\ee
Defining the vorticity
\be
\thalf \nabla \times \vec{v} = \vec{\omega}
\ee
we can express the Hamiltonian in terms of the orbital and spin angular momentum as
\be
H = \beta m + \vec{\alpha} \cdot {\bf p} - \vec{\omega} \cdot ({\bf L} + {\bf S}) \,.
\label{HLS}
\ee
It can also be written as
\be
H = \beta m + \vec{\alpha} \cdot {\bf p} - \vec{v} \cdot {\bf p} - \vec{\omega} \cdot {\bf S} \,.
\label{HpS}
\ee
This Hamiltonian is consistent with the literature.  When taking the nonrelativistic limit via the Foldy-Wouthuysen procedure, it is known that the orbital angular momentum term gives rise to the usual Coriolis and centrifugal forces \cite{Obukhov2013,Matsuo2017}.  The last term is the spin-rotation coupling.

The conserved current density is
\be
j^{\mu} = \bar{\psi} \hat{\gamma}^{\mu} \psi
\ee
which, as a 4-vector, should satisfy
\be
\partial_{\mu} j^{\mu} + \Gamma^{\nu}_{\nu \alpha} j^{\alpha} = 0 \,.
\ee
Since
\be
\Gamma^{\nu}_{\nu \alpha} = \frac{1}{\sqrt{-g}} \partial_{\alpha} \left(\sqrt{-g}\right)
\ee
where $g = {\rm det} \left( g_{\mu\nu} \right)$, and $g = -1$ in the present case, it follows that the current is ordinarily conserved.  One also finds by direct calculation from the Dirac equation that
\be
\partial_{\mu} j^{\mu} = 0 \,.
\ee

\section{Eigenvalues and Eigenvectors}
\label{SecIII}

As mentioned earlier, in this paper we are interested in the spin-rotation coupling.  The vorticity couples to the total angular momentum ${\bf J} = {\bf L} + {\bf S}$, and it is ${\bf J}$ which commutes with the Haniltonian of Eq. (\ref{HLS}) or (\ref{HpS}).  Nevertheless we shall henceforth drop the term $\vec{\omega} \cdot {\bf L} = \vec{v} \cdot {\bf p}$.  Because the vorticity in energy units is so small in high energy heavy ion collisions, typically on the order of several MeV, this appears justifiable.  Alternatively, one may restrict attention to the region near the origin where the orbital angular momentum is small and $|\vec{v}| \ll 1$.  Keeping the coupling of vorticity to orbital angular momentum complicates the problem significantly, and one should perhaps use an angular momentum basis rather than a momentum basis.

Consider the Hamiltonian $H = m \beta + \vec{\alpha} \cdot {\bf p} - \thalf \omega_0 \Sigma_3$.  Due to rotational symmetry around the $z$ axis we take the transverse momentum to be in the $x$ direction and set $p_2 = 0$.  Define $E = \sqrt{p^2 + m^2}$ and $E_3 = \sqrt{p_3^2 + m^2}$.  The two positive energy states have eigenvalues $E_{\pm} = \sqrt{ E^2 + \oneqt \omega_0^2 \pm \omega_0 E_3}$.  The unnormalized eigenvector for the upper sign is
\be
u_+ = \begin{pmatrix}
\dfrac{- p_1 p_3}{(E_3 + m) \left(E_+ + E_3 + \thalf \omega_0\right)} \\
\\
1 \\
\\
\dfrac{p_1}{E_+ + E_3 + \thalf \omega_0} \\
\\
\dfrac{-(E_3 - m)}{p_3} \\
\end{pmatrix} \,.
\ee 
and the unnormalized eigenvector for the lower sign is
\be
u_- = \begin{pmatrix}
1 \\
\\
\dfrac{p_3 (E_- - E_3 + \thalf \omega_0)}{p_1 (E_3 + m) } \\
\\
\dfrac{p_3}{E_3 + m} \\
\\
\dfrac{E_- - E_3 + \thalf \omega_0}{p_1} \\
\end{pmatrix} \,.
\ee 
These eigenvectors are orthogonal.  When $p_1 \rightarrow 0$, $u_+$ is an eigenstate of $\Sigma_3$ with eigenvalue $-1$ while $u_-$ is an eigenstate of $\Sigma_3$ with eigenvalue $+1$.

The two negative energy states have eigenvalues $-E_{\pm}$.  The unnormalized eigenvector for the upper sign is
\be
v_+ = \begin{pmatrix}
\dfrac{- \left( E_3 - m \right)}{p_3} \\
\\
\dfrac{- p_1}{E_+ + E_3 + \thalf \omega_0}\\
\\
1 \\
\\
\dfrac{p_1 \left( E_3 - m \right)}{p_3 \left(E_+ + E_3 + \thalf \omega_0\right)} \\
\end{pmatrix} \,.
\ee 
and the unnormalized eigenvector for the lower sign is
\be
v_- = \begin{pmatrix}
\dfrac{- \left( E_- - E_3 + \thalf \omega_0 \right)}{p_1}  \\
\\
\dfrac{p_3}{E_3 + m } \\
\\
\dfrac{- p_3 (E_- - E_3 + \thalf \omega_0)}{p_1 (E_3 + m) } \\
\\
1 \\
\end{pmatrix} \,.
\ee 
When $p_1 \rightarrow 0$, $v_+$ is an eigenstate of $\Sigma_3$ with eigenvalue $+1$ while $v_-$ is an eigenstate of $\Sigma_3$ with eigenvalue $-1$.

\section{Fluctuation Theory}
\label{SecIV}

The expressions given in Sec. \ref{SecII} remain valid no matter what direction in space the angular velocity is pointing in.  They also remain true if the vorticity is allowed to depend on $t$, although it cannot depend on space.  We write it as
\be
\vec{\omega}(t) = (\omega_1(t), \omega_2(t), \omega_0 + \omega_3(t))
\ee
where $\omega_0$ is the constant, average angular velocity and the $\omega_i(t)$ are small, fluctuating quantities whose averages are zero.  We want to calculate the equivalent of the Bloch equations for this situation, which entails using second order perturbation theory (second order response theory).  We mostly follow the notation of Ref. \cite{Schwabl_adv}.  See also Ref. \cite{Fabian}.  Note that the latter reference uses the density matrix formalism whereas we use the commutator formalism.

Consider a time independent Hamiltonian $H_0$.  The eigenvalues and eigenstates of $H_0$ are such that $H_0 |n\rangle = E_n |n\rangle$.  This is in the Heisenberg picture where
$|n\rangle \equiv |n\rangle_{\rm H} = e^{iH_0 t} |n,t\rangle_{\rm S}$ with the subscripts H and S referring to the Heisenberg and Schr\"odinger pictures.  Consider a time independent operator $A_{\rm S}$ in the Schr\"odinger picture.  In the Heisenberg picture it is $A_{\rm H}(t) =  e^{iH_0 t} A_{\rm S} \, e^{-iH_0 t}$.  The thermal average of this operator is
\ba
\langle A_{\rm H}(t) \rangle_0 &=& \frac{1}{Z_0} {\rm Tr} \left( e^{-\beta (H_0 - \mu N)} A_{\rm H}(t) \right) 
= \frac{1}{Z_0} {\rm Tr} \left( e^{-\beta (H_0 - \mu N)} A_S \right) \nonumber \\
&=& \frac{1}{Z_0} \sum_n e^{-\beta (E_n - \mu N_n)} \langle n| A_{\rm S} | n \rangle = \langle A_{\rm S} \rangle_0 \,.
\label{H0ensemble}
\ea
The subscript 0 on the right angular bracket indicates that the average is taken with respect to the Hamiltonian $H_0$ together with any conserved charge $N$.  The average is clearly time independent.  

Next consider the Hamiltonian $H(t) = H_0 + V(t)$, where $V(t)$ is a time dependent perturbation that vanishes when $t < 0$.  The time evolution operator for the full $H(t)$ is denoted by $U(t)$.  It satisfies the equation of motion
\be
\frac{d}{dt} U(t) = -i H(t) U(t) \,.
\ee
It can be factorized as
\be
U(t) = e^{-iH_0 t} U_{\rm I}(t) \,.
\ee
This leads to the equation of motion for $U_{\rm I}(t)$
\be
\frac{d}{dt} U_{\rm I}(t) = -i V_{\rm I}(t) U_{\rm I}(t)
\ee
where $V_{\rm I}(t) = e^{iH_0 t} V(t) \, e^{-iH_0 t}$ is the perturbation in the interaction picture.  This can be solved iteratively to yield
\be
U_{\rm I}(t) = 1 +\frac{1}{i} \int_0^t dt' V_{\rm I}(t') 
+ \frac{1}{i^2} \int_0^t dt' \int_0^{t'} dt'' V_{\rm I}(t') V_{\rm I}(t'') + \cdot\cdot\cdot \,.
\ee

The density matrix now becomes time dependent,
\be
\rho(t) = U(t) \rho_0 U^{\dagger}(t)  = e^{-iH_0 t} U_{\rm I}(t) \rho_0 U_{\rm I}^{\dagger}(t) e^{iH_0 t}\,,
\ee
and 
\be
\rho_{\rm I}(t)  = e^{iH_0 t} \rho(t) e^{-iH_0 t} = U_{\rm I}(t) \rho_0 U_{\rm I}^{\dagger}(t) \,,
\ee
where 
\be
\rho_0 = \frac{e^{-\beta (H_0 - \mu N)}}{Z_0} \,.
\ee
The $\rho_{\rm I}$ satisfies
\be
\frac{d\rho_{\rm I}}{dt} = -i \left[ V_{\rm I}, \rho_{\rm I} \right] \,.
\ee
Averages are taken with $\rho(t)$ instead of $\rho_0$.  This is denoted by replacing the subscript 0 with $t$ on the right angular bracket.  Thus
\be
\langle A \rangle_t = {\rm Tr} \left( \rho(t) A_{\rm S} \right) = {\rm Tr} \left( e^{-iH_0 t} \rho_{\rm I}(t) \, e^{iH_0 t} A_{\rm S} \right)
= {\rm Tr} \left( \rho_{\rm I}(t) A_{\rm I} \right) \,.
\ee
The average is representation independent, as it must be.

We are interested in the equation of motion for $\langle A \rangle_t$.  To this end we calculate
\be
\frac{d}{dt} \left[ \rho_0 U^{\dagger}(t) A_{\rm S} \, U(t) \right] = \frac{d}{dt} \left[ \rho_0 U_{\rm I}^{\dagger}(t) A_{\rm I}(t) \, U_{\rm I}(t) \right] 
= \rho_0 \left[ U_{\rm I}^{\dagger} \frac{dA_{\rm I}}{dt} U_{\rm I} + \frac{dU_{\rm I}^{\dagger}}{dt} A_{\rm I} U_{\rm I} + U_{\rm I}^{\dagger} A_{\rm I} \frac{dU_{\rm I}}{dt} \right]
\ee
and take the trace.  Using the equation of motion $dA_{\rm I}/dt = i [H_0, A_{\rm I}]$, the first term on the far RHS contributes to $d\langle A \rangle_t/dt$ the term
\be
i {\rm Tr} \left( \rho_0  U_{\rm I}^{\dagger}(t) [H_0, A_{\rm I}(t)] U_{\rm I}(t) \right) = i {\rm Tr} \left( \rho_{\rm I}(t) [H_0, A_{\rm I}(t)] \right) 
= i {\rm Tr} \left( \rho(t) [H_0, A_{\rm S}] \right) = i \langle  [H_0, A_{\rm S}] \rangle_t \,.
\ee
The second and third terms on the far RHS are
\bd
i \rho_0 \left[ V_{\rm I}(t), A_{\rm I}(t) \right] -\rho_0 \int_0^t dt' [ V_{\rm I}(t'),  [V_{\rm I}(t), A_{\rm I}(t) ]]
\ed
up to and including terms of second order in the perturbation.  Taking the trace yields
\be
\frac{d}{dt} \langle A \rangle_t =  i \langle  [H_0, A_{\rm S}] \rangle_t  + i \langle [ V_{\rm I}(t), A_{\rm I}(t)] \rangle_0
- \int_0^t dt' \langle [ V_{\rm I}(t'),  [V_{\rm I}(t), A_{\rm I}(t) ]] \rangle_0 \,.
\ee

Next we shall perform an ensemble or time average over the fluctuating fields.  We assume that $\overline{V(t)} = \overline{V_{\rm I}(t)} = 0$.  The average of the product 
$\overline{V(t)V(t')} = \overline{V_{\rm I}(t)V_{\rm I}(t')}$ is not zero but is assumed to be correlated on a time scale of $\tau_c$.  It is also assumed that fluctuations induced in 
$A_{\rm I}(t)$ are small enough that we may ignore $\overline{V_{\rm I}(t')A_{\rm I}(t) }$. This is coarse graining, also sometimes called the Born approximation.  Therefore, up to second order in the fluctuations we have
\be
\frac{d}{dt} \langle A \rangle_t =  i \langle  [H_0, A_{\rm S}] \rangle_t 
- \int_0^t dt' \langle \langle [ V_{\rm I}(t'),  [V_{\rm I}(t), A_{\rm I}(t) ]] \rangle \rangle_0 \,.
\label{2nd}
\ee
Here the double angular bracket means that averaging over the fluctuating fields is performed in addition to the thermal ensemble average of Eq. (\ref{H0ensemble}).  It is a more convenient notation than the overline.

\section{Massive Pauli Particle}
\label{SecV}

The Hamiltonian for a massive, nonrelativistic particle with spin one-half is
\be
H_0 = \frac{p^2}{2m} - \thalf \omega_0 \sigma_3 \,.
\ee
Since the kinetic energy commutes with the spin operator, this is basically the simple spin model presented in Section IV of Ref. \cite{Fabian}.  Let us apply the results of Sec. \ref{SecIV} with $H_0 = - \thalf \omega_0 \sigma_3$.  Then
\be
e^{ i H_0 t} = \cos\left( \thalf \omega_0 t \right) - i \sin\left( \thalf \omega_0 t \right) \sigma_3 \,.
\label{simples}
\ee
Recall the well known similarity transformations
\ba
\sigma_1(t) \equiv e^{ i H_0 t} \sigma_1 e^{ -i H_0 t} &=& \cos(\omega_0 t) \sigma_1 + \sin(\omega_0 t) \sigma_2 \nonumber \\
\sigma_2(t) \equiv e^{ i H_0 t} \sigma_2 e^{ -i H_0 t} &=& \cos(\omega_0 t) \sigma_2 - \sin(\omega_0 t) \sigma_1 \nonumber \\
\sigma_3(t) \equiv e^{ i H_0 t} \sigma_3 e^{ -i H_0 t} &=& \sigma_3 \,.
\ea
From here on in, whenever a Pauli or Dirac matrix appears without a time argument it is understood to remain unaffected by the time evolution.  Equivalently, it is evaluated at $t=0$.

We are most interested in the operator $A_S = \sigma_3$.  It quickly follows that
\be
[V_{\rm I}(t), \sigma_3] = i \omega_1(t) \sigma_2(t) -i  \omega_2(t) \sigma_1(t) \,.
\ee
The model assumes that fluctuations in different directions in Cartesian coordinates are uncorrelated, namely
\be
\overline{\omega_i(t) \omega_j(t')} = \overline{\omega_i^2} \, e^{-|t-t'|/\tau_c} \, \delta_{ij}
\label{wcorr}
\ee
where $\tau_c$ is a correlation time.  We can write the double commutator as
\be
[[ V_{\rm I}(t'),  [V_{\rm I}(t), \sigma_3 ]] = W_{11} + W_{22} + {\rm cross \, terms}
\ee
with
\ba
W_{11} &=& \omega_1(t) \omega_1(t') \cos[\omega_0 (t-t')] \sigma_3 \nonumber \\
W_{22} &=& \omega_2(t) \omega_2(t') \cos[\omega_0 (t-t')] \sigma_3 \,.
\ea
Cross terms involve $\omega_i(t) \omega_j(t')$ with $i \ne j$; these average to zero.  Averaging involves the integral
\be
\int_0^t dt' \, e^{-|t-t'|/\tau_c} \cos[\omega_0 (t-t')] \,.
\label{intica}
\ee
For $t \gg \tau_c$ this integral becomes
\be
T_0 = \frac{\tau_c}{1 + \omega_0^2 \tau_c^2} \,.
\label{inticb}
\ee
Recognizing that we are interested in the small departure from the equilibrium value of the $z$ component of the spin
 \be
s_{3}^{\rm eq} = \thalf \tanh(\beta \omega_0/2) 
\ee
we find that
\be
\frac{ds_3}{dt} = - \frac{s_3 - s_{3}^{\rm eq}}{\tau}
\label{depart}
\ee
where the relaxation time is given by
\be
\frac{1}{\tau} = \left( \overline{\omega_1^2} + \overline{\omega_2^2} \right) T_0 \,.
\ee
This agrees with the Eq. (IV.33) of Ref. \cite{Fabian} which based its calculations on the density matrix.

Analogous calculations can be done for the components of spin perpendicular to the vorticity.  The double communators needed for the $x$ component of spin are
\ba
[\sigma_2(t'), [\sigma_2(t), \sigma_1(t)]] &=& 4 \sigma_1(t') \nonumber \\
{[\sigma_3, [\sigma_3, \sigma_1(t)]]} &=& 4 \sigma_1(t) 
\ea
and for the $y$ component
\ba
[\sigma_1(t'), [\sigma_1(t), \sigma_2(t)]] &=& 4 \sigma_2(t') \nonumber \\
{[\sigma_3, [\sigma_3, \sigma_2(t)]]} &=& 4 \sigma_2(t) \,.
\ea
One encounters integrals like
\be
\int_0^t dt' e^{-(t-t')/\tau_c} \sigma_2(t') = T_0 \left[ \sigma_2(t) + \omega_0 \tau_c \sigma_1(t) \right] \,.
\ee
Putting them all together results in the remaining two spin equations.
\ba
\frac{ds_1}{dt} &=& \omega_0 \left( 1 + \overline{\omega_2^2} \tau_c T_0 \right) s_2 
- \left( \overline{\omega_3^2}\tau_c + \overline{\omega_2^2} T_0 \right) s_1 \nonumber \\
\frac{ds_2}{dt} &=& -\omega_0 \left( 1 + \overline{\omega_1^2} \tau_c T_0 \right) s_1 
- \left( \overline{\omega_3^2}\tau_c + \overline{\omega_1^2} T_0 \right) s_2
\ea
These are the same as Eqs. (IV.31) and (IV.32) of Ref. \cite{Fabian} apart from two points.  First, the sign of our $H$ is the opposite of theirs which flips the sign of the spin precession terms.  Second, the spin precession terms in Eqs. (IV.31) and (IV.32) have a correction to the spin precession frequency which is reduced (minus sign), whereas our result indicates an enhancement (plus sign).  This might be a misprint, or it might be traced to an incorrect reading of the sign of the imaginary part of Eq. (IV.22).  

\section{Massless Dirac Particle}
\label{SecVI} 

In this section we apply the general formulas to the case of  a massless Dirac particle.  We focus on the situation where the momentum is parallel to the vorticity.  The general case is much more involved and does not provide significantly more useful information.  

The Hamiltonian is
\be
H_0 = p \alpha_3 - \thalf \omega_0 \Sigma_3 \,.
\ee
Note that $\Sigma_3$ commutes with $H_0$.  The time evolution operator is (for details see the appendix)
\ba
e^{iH_0 t} &=& \cos \left(\thalf \omega_0 t \right) \cos (pt) \, I + \sin \left(\thalf \omega_0 t \right) \sin (pt) \, \gamma_5 \nonumber \\
&+&i \cos \left(\thalf \omega_0 t \right) \sin (pt) \, \alpha_3 - i \sin \left(\thalf \omega_0 t \right) \cos (pt) \, \Sigma_3 \,.
\ea
The similarity transformations of the $\Sigma$ matrices are
\ba
\Sigma_1(t) = e^{ i H_0 t} \Sigma_1 e^{ -i H_0 t} &=& \cos(\omega_0 t) \cos(2pt) \Sigma_1 
+ \sin(\omega_0 t) \cos(2pt) \Sigma_2 \nonumber \\
&+&\sin(\omega_0 t) \sin(2pt) \alpha_1 - \cos(\omega_0 t) \sin(2pt) \alpha_2 \nonumber \\
\Sigma_2(t) = e^{ i H_0 t} \Sigma_2 e^{ -i H_0 t} &=&  \cos(\omega_0 t) \cos(2pt)  \Sigma_2 
- \sin(\omega_0 t) \cos(2pt) \Sigma_1 \nonumber \\
&+& \sin(\omega_0 t) \sin(2pt) \alpha_2 + \cos(\omega_0 t) \sin(2pt) \alpha_1 \nonumber \\
\Sigma_3(t) = e^{ i H_0 t} \Sigma_3 e^{ -i H_0 t} &=& \Sigma_3 \,.
\ea
The single and double commutators needed for the fluctuations of $\omega_i(t)$ in the $i=1$ direction are
\be
\left[ \Sigma_1(t), \Sigma_3 \right] = -2i \Sigma_2(t)
\ee
and
\ba
\left[ \Sigma_1(t'), \left[ \Sigma_1(t), \Sigma_3 \right] \right] &=& 2 \left\{ \cos[(\omega_0 + 2p)(t-t')] + \cos[(\omega_0 - 2p)(t-t')] \right\} \Sigma_3 \nonumber \\
&+& 2 \left\{ \cos[(\omega_0 - 2p)(t-t')] - \cos[(\omega_0 + 2p)(t-t')] \right\} \alpha_3 \,.
\ea
The latter shows that $\langle \Sigma_3 \rangle$ and  $\langle \alpha_3 \rangle$ are coupled.  Therefore we also need
\ba
\left[ \Sigma_1(t), \alpha_3 \right] &=& 2i \sin(\omega_0 t) \cos(2pt) \alpha_1 - 2i \cos(\omega_0 t) \cos(2pt) \alpha_2 \nonumber \\
&-& 2i \cos(\omega_0 t) \sin(2pt) \Sigma_1 - 2i \sin(\omega_0 t) \sin(2pt) \Sigma_2 \nonumber \\
&=& -2i \alpha_2(t)
\ea
and
\ba
\left[ \Sigma_1(t'), \left[ \Sigma_1(t), \alpha_3 \right] \right] &=& 2 \left\{ \cos[(\omega_0 + 2p)(t-t')] + \cos[(\omega_0 - 2p)(t-t')] \right\} \alpha_3 \nonumber \\
&+& 2 \left\{ \cos[(\omega_0 - 2p)(t-t')] - \cos[(\omega_0 + 2p)(t-t')] \right\} \Sigma_3 \,.
\ea

Averaging over the fluctuating fields $\omega_i(t)$ can be performed using Eqs. (\ref{intica}) and (\ref{inticb}).  This results in
\ba
\frac{d}{dt} \langle \Sigma_3 \rangle_t &=& - \frac{\langle \Sigma_3 \rangle_t}{\tau_1} - \frac{\langle \alpha_3 \rangle_t}{\tau_2} \nonumber \\
\frac{d}{dt} \langle \alpha_3 \rangle_t &=& - \frac{\langle \alpha_3 \rangle_t}{\tau_1} - \frac{\langle \Sigma_3 \rangle_t}{\tau_2}
\label{coupledmassless}
\ea
where
\ba
\frac{1}{\tau_1} &=& \thalf \left( \overline{\omega_1^2} + \overline{\omega_2^2} \right) (T_- + T_+) \nonumber \\
\frac{1}{\tau_2} &=& \thalf \left( \overline{\omega_1^2} + \overline{\omega_2^2} \right) (T_- - T_+)
\ea
and
\be
T_{\pm} = \frac{\tau_c}{1 + (2p \pm \omega_0)^2 \tau_c^2} \,.
\ee
This makes use of the rotational symmetry around the vorticity axis.  The normal modes are
\be
\frac{d}{dt} \langle \Sigma_3 \pm \alpha_3 \rangle_t = - \frac{\langle \Sigma_3 \pm \alpha_3 \rangle_t}{\tau_{\pm}}
\label{depart2}
\ee
where
\be
\frac{1}{\tau_{\pm}} = \left( \overline{\omega_1^2} + \overline{\omega_2^2} \right) T_{\mp} \,.
\ee
In Eq. (\ref{depart2}) it is to be understood that these represent departures from the equilibrium values, otherwise the equilibrium values should be inserted by hand on the right hand side, as in Eq. (\ref{depart}).  Of phenomenological interest is the situation where $\omega_0 \ll |p|$.  In that limit
\be
\frac{1}{\tau_{\pm}} \approx \frac{ \left( \overline{\omega_1^2} + \overline{\omega_2^2} \right) \tau_c}
{1 + 4p^2 \tau_c^2} \,.
\ee

\section{Massive Dirac Particle}
\label{SecVII}

Now add a mass term but keep the momentum parallel to the vorticity.  The Hamiltonian is
\be
H_0 = m \beta + p \alpha_3 - \thalf \omega_0 \Sigma_3 \,.
\ee
The evolution operator has the form
\be
e^{iH_0 t} = C_1 I + C_2 \gamma_5 + C_3 \beta \Sigma_3 + iC_4 \beta + i C_5 \alpha_3 + i C_6 \Sigma_3 \,.
\ee
The coefficients are calculated in the appendix with the result
\ba
C_1 &=&  \cos(Et) \cos\left(\thalf \omega_0 t \right) \nonumber \\
C_2 &=&  \frac{p}{E} \sin(Et) \sin\left(\thalf \omega_0 t \right) \nonumber \\
C_3 &=&  \frac{m}{E} \sin(Et) \sin\left(\thalf \omega_0 t \right) \nonumber \\
C_4 &=&  \frac{m}{E} \sin(Et) \cos\left(\thalf \omega_0 t \right) \nonumber \\
C_5 &=&  \frac{p}{E} \sin(Et) \cos\left(\thalf \omega_0 t \right) \nonumber \\
C_6 &=&  - \cos(Et) \sin\left(\thalf \omega_0 t \right) \,.
\label{massiveCs}
\ea

Operators in the interaction picture can be obtained from those in the Schr\"odinger picture with tedious algebra.  Of particular interest are
\be
\Sigma_1(t) = B_1(t) \Sigma_1 + B_2(t) \Sigma_2
+ B_3(t) \alpha_1 - B_4(t) \alpha_2 + i B_5(t) \beta \alpha_1 + i B_6(t) \beta \alpha_2 \,,
\ee
\be
\Sigma_2(t) = B_1(t) \Sigma_2 - B_2(t) \Sigma_1
+ B_3(t) \alpha_2 + B_4(t) \alpha_1 + i B_5(t) \beta \alpha_2 - i B_6(t) \beta \alpha_1 \,,
\ee 
while $\Sigma_3(t) = \Sigma_3$ on account of the fact that it commutes with $H_0$.  The $B_i$ are given in the appendix.  We shall also need
\be
\alpha_3(t) =  \left[ \frac{p^2}{E^2} + \frac{m^2}{E^2}  \cos(2Et) \right] \alpha_3 + \frac{mp}{E^2} \left[ 1 - \cos(2Et) \right] \beta
+ i \frac{m}{E} \sin(2Et) \beta \alpha_3 \,,
\ee
and
\be
\beta(t) = \left[ \frac{m^2}{E^2} + \frac{p^2}{E^2} \cos(2Et) \right] \beta + \frac{mp}{E^2} \left[ 1 - \cos(2Et) \right] \alpha_3
- i \frac{p}{E} \sin(2Et) \beta \alpha_3 \,.
\ee
One can check that $m \beta(t) + p \alpha_3(t) = m \beta + p \alpha_3$, as it should be since this operator commutes with $H_0$.  From these the final relevant one is easily found to be
\be
\beta(t) \alpha_3(t) = \cos(2Et) \beta \alpha_3 - i \frac{p}{E} \sin(2Et) \beta + i \frac{m}{E} \sin(2Et) \alpha_3 \,.
\ee

What is needed for fluctuations for $\Sigma_3$ are the double commutators
\ba
\left[ \Sigma_1(t'), \left[ \Sigma_1(t), \Sigma_3 \right] \right] &=& \left[ \Sigma_2(t'), \left[ \Sigma_2(t), \Sigma_3 \right] \right] \nonumber\\
\left[ \Sigma_1(t'), \left[ \Sigma_1(t), \alpha_3 \right] \right] &=& \left[ \Sigma_2(t'), \left[ \Sigma_2(t), \alpha_3 \right] \right] \nonumber \\
\left[ \Sigma_1(t'), \left[ \Sigma_1(t), \beta \right] \right] &=& \left[ \Sigma_2(t'), \left[ \Sigma_2(t), \beta \right] \right]  \nonumber \\
\left[ \Sigma_1(t'), \left[ \Sigma_1(t), \beta \alpha_3 \right] \right] &=& \left[ \Sigma_2(t'), \left[ \Sigma_2(t), \beta \alpha_3 \right] \right] \,.
\ea
Explicit expressions in terms of the products $B_i(t') B_j(t)$ are given in the appendix.  After a fair amount of algebra these lengthy expressions can be put into a more useful form by defining $\theta_{\pm} = (2E \pm \omega_0)(t-t'))$ and $\theta_0 = \omega_0 (t-t')$.
\ba
\lefteqn{ \left[ \Sigma_1(t'), \left[ \Sigma_1(t), \Sigma_3 \right] \right] = 2\Bigg[ \frac{p^2}{E^2} \big( \cos\theta_- + \cos\theta_+ \big) 
+ 2 \frac{m^2}{E^2} \cos\theta_0  \Bigg] \Sigma_3} \nonumber \\
&+& 2 \frac{p}{E} \Bigg[ \big( \cos\theta_- - \cos\theta_+ \big) \Bigg( \frac{p^2}{E^2} +  \frac{m^2}{E^2} \cos(2Et) \Bigg) 
+ \frac{m^2}{E^2} \big( 2 \sin\theta_0 +\sin\theta_- - \sin\theta_+ \big) \sin(2Et) \Bigg] \alpha_3\nonumber \\
&+& 2 \frac{m p^2}{E^3} \Bigg[ \big( \cos\theta_- - \cos\theta_+ \big) \big( 1 - \cos(2Et) \big)
-  \big( 2 \sin\theta_0 + \sin\theta_- - \sin\theta_+ \big) \sin(2Et)  \Bigg] \beta \nonumber \\
&+& 2i \frac{mp}{E^2} \Bigg[ \big( \cos\theta_- - \cos\theta_+ \big) \sin(2Et)  - \big( 2 \sin\theta_0 +\sin\theta_- - \sin\theta_+  \big) \cos(2Et) \Bigg] \beta \alpha_3
\ea
\ba
\lefteqn{ \left[ \Sigma_1(t'), \left[ \Sigma_1(t), \alpha_3 \right] \right] = 
2 \frac{p}{E} \bigg[ \cos\theta_- - \cos\theta_+ \bigg]  \Sigma_3} \nonumber\\
&+& 2 \bigg[ \frac{p^2}{E^2} \big( \cos\theta_- + \cos\theta_+ \big) + 2\frac{m^2}{E^2} \cos\theta_0 \cos (2Et) \bigg] \alpha_3 \nonumber\\
&+& 2\frac{m p}{E^2} \bigg[ \cos\theta_- + \cos\theta_+ - 2 \cos\theta_0 \cos (2 E t) \bigg] \beta \nonumber \\
&+& 4i \frac{m}{E} \bigg[ \cos\theta_0 \sin (2 E t)  \bigg] \beta \alpha_3 
\ea
\ba
\left[ \Sigma_1(t'), \left[ \Sigma_1(t), \beta \right] \right] &=& 0
\ea
\ba
\lefteqn{ \left[ \Sigma_1(t'), \left[ \Sigma_1(t), \beta \alpha_3 \right] \right] = 2 i \frac{m p}{E^2} \bigg[  2 \sin\theta_0 + \sin\theta_- - \sin\theta_+  \bigg] \Sigma_3} \nonumber\\
&+& 2 i \frac{m}{E} \bigg[  \bigg( 2\frac{m^2}{E^2} \cos\theta_0 + \frac{p^2}{E^2} \big( \cos\theta_- + \cos\theta_+ \big) \bigg) \sin(2Et) \nonumber \\
&+& \frac{p^2}{E^2} \big( \sin\theta_- + \sin\theta_+ \big) \big(1 - \cos(2Et) \big) \bigg] \alpha_3 \nonumber\\
&+& 2 i \frac{p}{E} \bigg[ \big( \sin\theta_- + \sin\theta_+ \big) \Big( \frac{m^2}{E^2} + \frac{p^2}{E^2} \cos(2Et) \Big)
- \bigg( 2 \frac{m^2}{E^2} \cos\theta_0 + \frac{p^2}{E^2} \big( \cos\theta_- + \cos\theta_+ \big) \bigg) \sin(2Et) \bigg] \beta \nonumber\\
&+& 2 \bigg[ \bigg( 2  \frac{m^2}{E^2} \cos\theta_0 + \frac{p^2}{E^2} \big( \cos\theta_- + \cos\theta_+ \big) \bigg) \cos (2 E t) 
+ \frac{p^2}{E^2} \big( \sin\theta_- + \sin\theta_+ \big) \sin(2Et) \bigg] \beta \alpha_3 
\ea

The following integrals are useful for averaging over the fluctuations.  The arrows represent the steady state where $\tau_c \ll t$.
\be
\int_0^t dt' \, e^{-(t-t')/\tau_c} \cos[\omega_0 (t-t')] \rightarrow T_0 \,.
\ee
\be
\int_0^t dt' \, e^{-(t-t')/\tau_c} \sin[\omega_0 (t-t')] \rightarrow \omega_0 \tau_c T_0 \,.
\ee
\be
\int_0^t dt' e^{-(t-t')/\tau_c} \cos[ (2E \pm \omega_0)(t-t')] \rightarrow T_{\pm} \
\label{cosave}
\ee
\be
\int_0^t dt' e^{-(t-t')/\tau_c} \sin[ (2E \pm \omega_0)(t-t')] \rightarrow (2E \pm \omega_0) \tau_c T_{\pm} 
\label{sinave}
\ee
Here
\be
T_{\pm} = \frac{\tau_c}{1 + (2E \pm \omega_0)^2 \tau_c^2}
\ee
and
\be
T_0 = \frac{\tau_c}{1 + \omega_0^2 \tau_c^2}
\ee
as defined earlier.

The results of performing the integration over $t'$ in the steady state are
\ba
\lefteqn{ \int_0^t dt' e^{-(t-t')/\tau_c} \left[ \Sigma_1(t'), \left[ \Sigma_1(t), \Sigma_3 \right] \right] = 2\Bigg[ \frac{p^2}{E^2} \big( T_- + T_+ \big) 
+ 2 \frac{m^2}{E^2} T_0  \Bigg] \Sigma_3} \nonumber \\
&+& 2 \frac{p}{E} \Bigg[ \big( T_- - T_+ \big) \Bigg( \frac{p^2}{E^2} +  \frac{m^2}{E^2} \cos(2Et) \Bigg) 
+ \tau_c \frac{m^2}{E^2} \Bigg( 2 \omega_0 T_0 + (2E-\omega_0)T_-  - (2E+\omega_0) T_+\Bigg) \sin(2Et) \Bigg] \alpha_3 \nonumber \\
&+& 2 \frac{m p^2}{E^3} \Bigg[ \big( T_- - T_+ \big) \big( 1 - \cos(2Et)\big) 
- \tau_c \Bigg( 2 \omega_0 T_0 + (2E-\omega_0)T_-  - (2E+\omega_0) T_+\Bigg) \sin(2Et)  \Bigg] \beta \nonumber \\
&+& 2i \frac{mp}{E^2} \Bigg[ \big( T_- - T_+ \big) \sin(2Et)  
- \tau_c \Bigg( 2 \omega_0 T_0 + (2E-\omega_0)T_-  - (2E+\omega_0) T_+\Bigg) \cos(2Et) \Bigg] \beta \alpha_3
\ea

\ba
\lefteqn{ \int_0^t dt' e^{-(t-t')/\tau_c} \left[ \Sigma_1(t'), \left[ \Sigma_1(t), \alpha_3 \right] \right] = 
2 \frac{p}{E} \bigg[ T_- - T_+ \bigg]  \Sigma_3} \nonumber\\
&+& 2 \bigg[ \frac{p^2}{E^2} \big( T_- + T_+ \big) + 2\frac{m^2}{E^2} T_0 \cos (2Et) \bigg] \alpha_3 \nonumber\\
&+& 2\frac{m p}{E^2} \bigg[ T_- + T_+ - 2 T_0 \cos (2 E t) \bigg] \beta \nonumber \\
&+& 4i \frac{m}{E} \bigg[ T_0 \sin (2 E t) \bigg] \beta \alpha_3 
\ea

\ba
\lefteqn{ \int_0^t dt' e^{-(t-t')/\tau_c} \left[ \Sigma_1(t'), \left[ \Sigma_1(t), \beta \alpha_3 \right] \right] = 2 i \tau_c \frac{m p}{E^2} 
\bigg[  2 \omega_0 T_0 + (2E - \omega_0) T_- - (2E + \omega_0) T_+  \bigg] \Sigma_3} \nonumber\\
&+& 2 i \frac{m}{E} \bigg[  \bigg( 2\frac{m^2}{E^2} T_0 + \frac{p^2}{E^2} \big( T_- + T_+ \big) \bigg) \sin(2Et) \nonumber \\
&+& \tau_c \frac{p^2}{E^2} \big( (2E - \omega_0) T_- + (2E + \omega_0) T_+ \big) \big(1 - \cos(2Et) \big) \bigg] \alpha_3 \nonumber\\
&+& 2 i \frac{p}{E} \bigg[ \tau_c \big( (2E - \omega_0) T_- + (2E + \omega_0) T_+ \big) \Big( \frac{m^2}{E^2} + \frac{p^2}{E^2} \cos(2Et) \Big) \nonumber \\
&-& \bigg( 2 \frac{m^2}{E^2} T_0 + \frac{p^2}{E^2} \big( T_- + T_+ \big) \bigg) \sin(2Et) \bigg] \beta \nonumber\\
&+& 2 \bigg[ \bigg( 2  \frac{m^2}{E^2} T_0 + \frac{p^2}{E^2} \big( T_- + T_+ \big) \bigg) \cos (2 E t) \nonumber \\
&+& \tau_c \frac{p^2}{E^2} \big( (2E - \omega_0) T_- + (2E + \omega_0) T_+ \big) \sin(2Et) \bigg] \beta \alpha_3 
\ea
These still need to be expressed in terms of the time-dependent operators in the Dirac basis.  The relationships are given in the appendix.

Finally, we need the commutators of the operators in the interaction picture with the unperturbed Hamiltonian, which are
\ba
\left[ H_0 , \Sigma_3(t) \right] &=& 0 \nonumber \\
\left[ H_0 , \alpha_3(t) \right] &=& 2 m \beta \alpha_3 (t) \nonumber \\
\left[ H_0 , \beta(t) \right] &=&  -2 p \beta \alpha_3 (t)  \nonumber \\
\left[ H_0 , \beta(t) \alpha_3(t) \right] &=&  2m \alpha_3 (t) -2 p \beta (t) \,.
\ea

Then the equations of motion can be written in matrix form as
\be
\begin{split}
 \frac{d}{dt} \begin{pmatrix}   \langle \Sigma_3(t) \rangle \\  
                       \langle \alpha_3(t) \rangle \\
                       \langle i\beta(t) \alpha_3(t) \rangle \\
                       \langle \beta(t) \rangle
      \end{pmatrix} =&       
      \begin{pmatrix} 
      -h_0 && -h_2 && h_1 && 0\\
      -h_2&& -h_0 && 2m && -h_3\\
      h_1 && -2m &&-h_0 &&  (2p+h_4) \\
      0 && 0 && -2p && 0
      \end{pmatrix}     
      \begin{pmatrix}   \langle \Sigma_3(t) \rangle \\  
                       \langle \alpha_3(t) \rangle \\
                       \langle i\beta(t) \alpha_3(t) \rangle \\
                       \langle \beta(t) \rangle
      \end{pmatrix}
\end{split}
\ee
where
\be
\begin{split}
 h_0 =& \left[ \frac{p^2}{2E^2} \Big (T_- +T_+ \Big) + \frac{m^2}{E^2} T_0 \right]  \overline{\omega_{\perp}^2} \\
 h_1=& \frac{\tau_c mp}{2 E^2} \Big[ 2 \omega_0 T_0 + (2 E -\omega_0)T_- - (2 E+\omega_0)T_+ \Big]  \, \overline{\omega_{\perp}^2} \\
 h_2 =& \frac{p}{2 E} \Big( T_- - T_+ \Big) \, \overline{\omega_{\perp}^2} \\
 h_3=& \frac{m p}{2 E^2} \Big( T_- + T_+ -2 T_0 \Big) \, \overline{\omega_{\perp}^2} \\
 h_4=& \frac{\tau_c p}{2E} \Big[ (2E - \omega_0) T_- + (2E + \omega_0) T_+ \Big] \, \overline{\omega_{\perp}^2}
\end{split}
\ee
with
\be
\overline{\omega_{\perp}^2} = \overline{\omega_1^2} + \overline{\omega_2^2} \,.
\ee
As in the case of massless Dirac particles, it is understood that we are solving for the departures from the equilibrium values.

The eigenvalues $\lambda$ are found from a fourth order polynomial.  Defining $\lambda' = \lambda + h_0$ for convenience this polynomial is
\ba
P &=& \lambda'^4 - h_0 \lambda'^3 +\left( 4E^2 + 2ph_4 - h_1^2 - h_2^2 \right) \lambda'^2 \nonumber \\
&+& \left[ 4mph_3 + h_0 \left( h_1^2 + h_2^2 -4m^2 \right) \right] \lambda'
- 2ph_2 \left( 2ph_2 +h_2 h_4 - h_1 h_3 \right) \,.\
\ea
Consider some limiting cases.  

When $p=0$ then $h_1 = h_2 = h_3 = h_4 = 0$.  There is one zero eigenvalue belonging to $\langle \beta(t) \rangle$. The spin $\langle \Sigma_3(t) \rangle$ has eigenvalue 
$-\overline{\omega_{\perp}^2} T_0$.  The quantities $\langle \alpha_3(t) \rangle$ and $\langle i\beta(t) \alpha_3(t) \rangle$ are coupled with complex eigenvalues 
$-\overline{\omega_{\perp}^2} T_0 \pm 2 m i$.  There is only one relaxation time and it is the same as found earlier.

When $m=0$ then $h_1 = h_3 = 0$.  Defining $\lambda_{\pm} = \overline{\omega_{\perp}^2} T_{\mp}$, with $\lambda_+ \ge \lambda_-$, the remaining $h$'s are
\ba
h_0 &=& \thalf \left( \lambda_+ + \lambda_- \right) \nonumber \\
h_2 &=& \thalf \left( \lambda_+ - \lambda_- \right) \nonumber \\
h_4 &=& p \tau_c \left( \lambda_+ + \lambda_- \right) - \thalf \omega_0 \tau_c \left( \lambda_+ - \lambda_- \right) \,.
\ea
As we saw earlier, $\langle \Sigma_3(t) \rangle$ and $\langle \alpha_3(t) \rangle$ are coupled with eigenvalues $-\lambda_+$ and $-\lambda_-$.  The quantities $\langle \beta(t) \rangle$ and $\langle i\beta(t) \alpha_3(t) \rangle$ are coupled with eigenvalues
\bd
- \oneqt \left( \lambda_+ + \lambda_- \right) \pm \oneqt \sqrt{ \left( \lambda_+ + \lambda_- \right)^2 - 32 p \left( 2p + h_4 \right) } \,.
\ed
These are real for small momentum and become complex at larger momentum.  This momentum scale is essentially $(\lambda_+ + \lambda_-)/8$.  Then the pair of conjugate eigenvalues at larger momenta are to good approximation just
\bd
- \oneqt \left( \lambda_+ + \lambda_- \right) \pm 2 p i \,.
\ed
  
\section{Strange Quark Helicity Flip in Quark Gluon Plasma}
\label{SecVIII}

In this section we explore another mechanism for the relaxation rate for strange quark spin, which is spin/helicity flip in collisions of strange quarks with up or down quarks, antiquarks, or gluons.  As is well-known, the helicity of a massless quark is conserved in such collisions due to the vector coupling to gluons.  For a quark whose mass is small compared to its energy the cross section for helicity flip is proportional to $m^2$.  The current quark mass of the strange quark is around 100 MeV, whereas the temperature of the plasma might range from 500 MeV down to 200 MeV.  Therefore we may use this as an approximation to estimate the rate of helicity flip in the plasma.

A common approximation is the energy-dependent relaxation time approximation.  Consider the reaction $a+b \rightarrow c+d$.  The relaxation time $\tau_a(E_a)$ for species $a$ with energy $E_a$ as measured in the rest frame of the plasma is given by \cite{ChakrabortyKapusta2011,AlbrightKapusta2016}
\be
 \frac{1+d_a f_a^{\rm eq}}{\tau_a(E_a)}  =  \sum_{bcd} \frac{{\cal N}}{1+\delta_{ab}}
 \int d\Gamma_b \, d\Gamma_c \, d\Gamma_d \,  W(a,b|c,d) f_b^{\rm eq}  \left( 1 + d_c f_c^{\rm eq} \right) \left( 1 + d_d f_d^{\rm eq} \right)  \,.
 \label{eq:RTA:relaxationtimeQ}
\ee
Here the $f_i^{\rm eq}$ are Fermi-Dirac or Bose-Einstein distribution functions, and the $d_i$ are $-1$ for fermions and $+1$ for bosons.  The integration over phase space involves
\be
d\Gamma_i = \frac{d^3p_i}{(2\pi)^3} \,.
\ee
The $W$ is related to the dimensionless amplitude ${\cal M}$ by
\be
W(a,b|c,d) = \frac{(2\pi)^4 \delta^4\left( p_a+p_b-p_c-p_d\right)}
{2E_a2E_b2E_c2E_d} |{\cal M}(a,b|c,d)|^2 \, .
\ee
The $|{\cal M}(a,b|c,d)|^2$ is averaged over spin in both the initial and final states; this compensates the spin factor $2s_i+1$ in the phase space integration.
It is further related to the differential cross section by
\be
\frac{d\sigma}{dt} = \frac{1}{64\pi s} \frac{1}{p_*^2} |{\cal M}|^2
\label{xsection}
\ee
where $p_*$ is the initial state momentum in the center-of-momentum frame.  Finally, the ${\cal N}$ is a degeneracy factor for spin, color, and any other internal degrees of freedom.  Its value depends on how these variables are summed or averaged over in $|{\cal M}|^2$.  

In order to obtain analytical results we shall drop the Pauli suppression and Bose enhancement factors in the final state.  For the reactions $s+q \rightarrow s+q$ and 
$s+\bar{q} \rightarrow s+\bar{q}$, where $q = u$ or $d$, this will only slightly enhance the rate.  For the reaction $s+g \rightarrow s+g$ these two final state effects should approximately cancel.  Therefore we expect this approximation to result in an accurate or slight overestimate of the rate.

In Eq. (\ref{eq:RTA:relaxationtimeQ}) let $a$ be the incoming strange quark with momentum $p^{\mu}$, $b$ be the incoming light quark, anti-quark, or gluon with momentum $p^{\mu}_2$, $c$ be the outgoing strange quark with momentum $p^{\mu}_3$, and $d$ be the outgoing light quark, anti-quark, or gluon with momentum $p^{\mu}_4$.
With no loss of generality we will work in the rest frame of the plasma and take
\ba
{\bf \hat{p}} &=& (1, 0, 0) \nonumber \\
{\bf \hat{p}}_3 &=& (\cos\phi_3, \sin\phi_3, 0) \nonumber \\
{\bf \hat{p}}_4 &=& (\cos\phi_4 \sin\theta_4, \sin\phi_4 \sin\theta_4, \cos\theta_4) \,.
\ea
The Mandelstam variables are $s = (p+p_2)^2 = (p_3+p_4)^2$ and $t = (p_3-p)^2 = (p_4-p_2)^2$.  Following Ref. \cite{KapustaLichard1991} we insert integrations over $s$ and $t$ with a Dirac $\delta$-function for each.  This is a natural thing to do since $|{\cal M}|^2$ depends only on those two variables.

We can use the 3-dimensional $\delta$-function in Eq. (\ref{eq:RTA:relaxationtimeQ}) to eliminate the integration over ${\bf p}_2$.  Then consider the integral
\be
J = \int d\Omega_3 \, d\Omega_4 \, \delta\big(E + E_2 - E_3 - E_4\big) \, \delta\big(s - 2 E_3 E_4 (1 - {\bf \hat{p}}_3 \cdot {\bf \hat{p}}_4)\big)
\, \delta\big(t + 2 E E_3 (1 - {\bf \hat{p}} \cdot {\bf \hat{p}}_3)\big) \,,
\ee
where $E_2 = |{\bf p}_3 + {\bf p}_4 - {\bf p}|$.  In these coordinates
\bd
\int d\Omega_3 \, d\Omega_4 = \int_0^{2\pi} d\theta_3 \int_0^{\pi} d\phi_3 \, \sin\phi_3 
\int_0^{\pi} d\theta_4 \, \sin\theta_4 \int_0^{2\pi} d\phi_4  \,.
\ed
The result is
\be
J = \frac{2\pi E_2}{E E_3 E_4} \frac{\theta(y - E)}{\sqrt{Ax^2 + Bx + C}}
\ee
where
\ba
x &=& E_3 - E_4 \nonumber \\
y &=& E_3 + E_4 \nonumber \\
A &=& - \oneqt s^2 \nonumber \\
B &=& \thalf s(s+2t)(2E-y) \nonumber \\
C &=& st(s+t) + s^2 E (y-E) - \oneqt (s+2t)^2 y^2 \,.
\ea
Part of the remaining integration is $dE_3 dE_4 = \thalf dx dy$.  Integration over $x$ can be done immediately
\be
\int_{x_+}^{x_-} \frac{dx}{\sqrt{Ax^2 + Bx + C}} = \frac{2\pi}{s}
\ee
where
\bd
x_{\pm} = \frac{-B \pm \sqrt{B^2 - 4AC}}{2A}
\ed
since $B^2 - 4AC \ge 0$ and $A < 0$.  The integral over $y$ is simply
\be
\int_{E + s/4E}^{\infty} dy \, f_2^{\rm eq}(y-E) = T \ln \left( 1 \pm e^{-s/4ET} \right)^{\pm 1}
\ee
where the upper sign is for fermions and the lower sign for bosons.  The expression  (\ref{eq:RTA:relaxationtimeQ}) reduces to
\be
\frac{1 - f^{\rm eq}(E)}{\tau(E)}  = \frac{ {\cal N} \, T}{32 (2\pi)^3 E^2} \int \frac{ds}{s} \ln \left( 1 \pm e^{-s/4ET} \right)^{\pm 1} \int dt \, |{\cal M}(s,t)|^2 \,.
\ee

\begin{figure}[hh]
\centering
\includegraphics[scale=0.6]{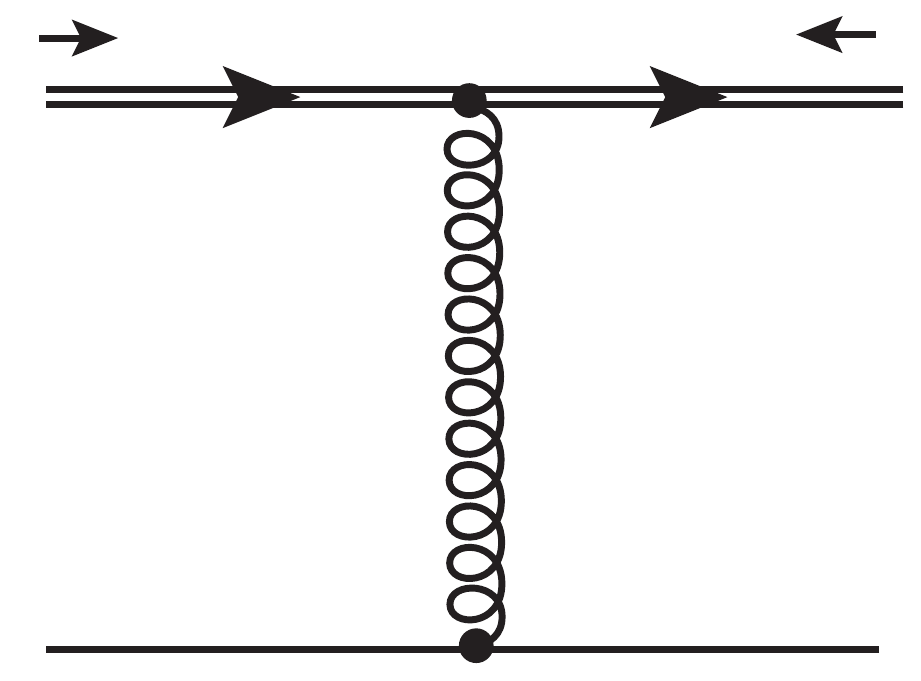}
\caption{Scattering of a massive strange quark (double line) which flips its helicity.  The massless quark or anti-quark cannot change its helicity. }
\label{qq}
\end{figure}

First consider the reaction $s+q \rightarrow s+q$ shown in Fig. \ref{qq}.  
The strange quark is allowed to flip its helicity because it has a small mass while the lighter quark is not.  There are four combinations: a positive helicity strange quark can flip to negative helicity with a light quark of either helicity, and a negative helicity strange quark can flip to positive helicity with a light quark of either helicity.  A straighforward calculation (see Appendix D) yields
\be
|{\cal M}|_{{\rm helicity \, flip}}^2 = - \frac{8}{9} g^4 m^2 \left[ \frac{1}{t} + \frac{s}{(s-m^2)^2} \right]
\label{Mflip} 
\ee
for any one of these four combinations.  Here $g$ is the QCD coupling constant.  In this expression initial colors are averaged over while final colors are summed over.  If we want the rate for helicity flip of the strange quark we should use the above expression along with ${\cal N} = 6$ because of scattering from a light quark of either helicity and of any of three colors.  The integral over $t$ is essentially the cross section.  Since a massless particle, the gluon, is being exchanged the cross section would be logarithmically divergent.  Many-body effects are necessary to screen this divergence.  

Here we follow Ref. \cite{KapustaLichard1991}.  We remove the region of phase space causing the divergence.  We integrate over
\be
-\frac{(s-m^2)^2}{s} + k_c^2 \le t \le -k_c^2 \,,
\ee
and similarly for $u$, where $k_c^2$ is an infrared cutoff.  This way of regulating the divergence treats $t$ and $u$ symmetrically.  It maintains $s + t + u = 2m^2$.  In order that the restricted range of $t$ be sensible means that a small region of $s$ must also be removed.  We should take $s \ge s_0$ where
\be
\frac{(s_0-m^2)^2}{s_0} = 2k_c^2 \,,
\ee
or $s_0 = m^2 + k_c^2 + k_c \sqrt{2m^2 + k_c^2}$.  For the photon production processes addressed in Ref. \cite{KapustaLichard1991} all quarks were massless so then $s_0 = 2k_c^2$.  The contribution from the region of phase space removed must be added back in using the method of hard thermal loops, as Ref. \cite{KapustaLichard1991} did.  Now the $t$ integration gives
\be
\int_{-(s-m^2)^2/s +k_c^2}^{-k_c^2} dt \, |{\cal M}|_{{\rm helicity \, flip}}^2 = \frac{8}{9} g^4 m^2 
\left\{ \ln\left[ \frac{(s - m^2)^2}{s k_c^2} -1 \right]  - 1 + \frac{2 s \, k_c^2}{(s-m^2)^2} \right\} \,.
\label{t-int}
\ee
The remaining integral over $s$ is
\be
I_q = 4 \int_{s_0}^{\infty} \frac{ds}{s} \left\{ \ln\left[ \frac{(s - m^2)^2}{s k_c^2} -1 \right]  - 1 + \frac{2 s \, k_c^2}{(s-m^2)^2} \right\} \ln \left( 1 + e^{-s/4ET} \right) \,.
\label{s-int1}
\ee
In Ref. \cite{KapustaLichard1991} it was shown that by adding the hard thermal loop contribution (the region of phase space removed) essentially replaced $k_c^2$ with the effective mass of the exchanged quark in the plasma, as defined not at zero momentum but at high momentum, such that $k_c^2 \propto g^2T^2$.  It is worth noting that the hard thermal loop contribution came from a branch cut, not a pole.  In the present case we would expect $k_c^2$ to be replaced by the effective mass of the exchanged gluon, $k_c^2 = m_P^2$, with
\be
m_P^2 = \ones g^2 \left( N_c + \thalf N_f \right) T^2
\ee
where $N_c$ is the number of colors and $N_f$ is the number of light flavors.  Performing the hard thermal loop calculation to validate this is outside the scope of this paper.

Allowing for scattering from $u$, $\bar u$, $d$ and $\bar d$ provides another factor of 4.  Putting it all together yields
\be
\frac{1}{\tau_q(E)}  = \frac{\alpha_s^2 \, T}{3 \pi} \frac{m^2}{E^2} \left( 1 + e^{-E/T} \right) I_q \,.
\ee
Unfortunately the expression for $I_q$ cannot be evaluated analytically. 

There is another way to approach this problem, which is to insert a static color electric screening mass in the gluon propagator.  Replace Eq. (\ref{Mflip}) with
\be
|{\cal M}|_{{\rm helicity \, flip}}^2 = - \frac{2}{9} g^4 m^2 \left[ t + \frac{s t^2}{(s-m^2)^2} \right] \frac{1}{(t - m_{\rm el}^2)^2}
\label{modMflip} 
\ee
where $m_{\rm el}^2 = 2 m_P^2$.  This is the approach used in the parton cascade model ZPC \cite{ZPC} as implemented in the AMPT model which simulates high energy heavy ion collisions \cite{AMPT}.  Although appealing, this approach is not manifestly gauge invariant so we do not pursue it here.

\begin{figure}[hh]
\centering
\includegraphics[scale=0.6]{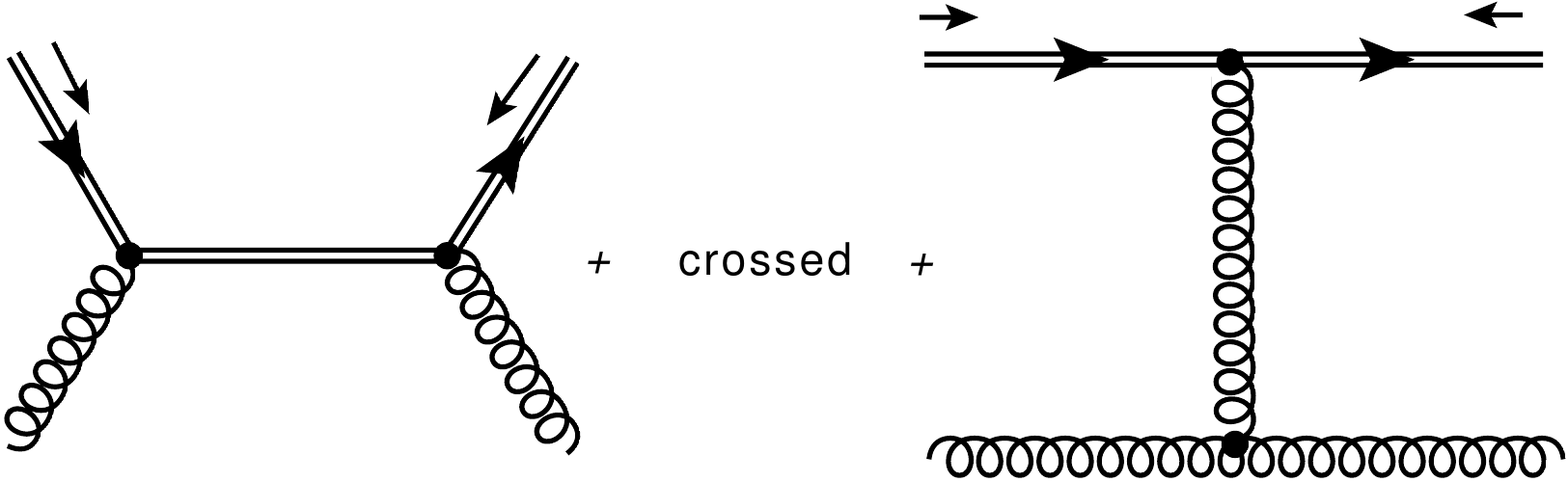}
\caption{Scattering of a massive strange quark (double line), which flips its helicity, with a gluon.}
\label{qg}
\end{figure}

Next consider the reaction $s+g \rightarrow s+g$ shown in Fig. \ref{qg}.
A lengthy calculation (see Appendix D) yields
\ba
|{\cal M}|^2_{\rm helicity \, flip} &=& \frac{4 g^4 m^2 (-t)}{3 s_m^4} \Bigg\{ 4 s^2 \left[ 3 + \frac{3 t}{u_m} + \frac{4 t^2}{3 u_m^3} \right] + 
8 (m^4 - s u)^2 \left[ \frac{3}{t^2} - \frac{3}{u_m t} + \frac{4}{3 u_m^2} \right] \nonumber \\
&+& \frac{7 m^4 t^2}{3 u_m^2} + 3 \left[ \frac{(m^4 - s u - s_m u_m)}{u_m} - \frac{2(m^4 - s u)}{t} \right]^2 \Bigg\} \,.
\ea
Here $s_m = s - m^2$, $u_m = u - m^2$, and $r^2 = m^4 - s u$.  Then $s_m + t + u_m = 0$.  This is obtained by summing over all colors in the initial and final states, and dividing by 3 to obtain the rate for scattering of a strange quark with a specified (average) color.  When integrating over $t$ or $u$ it is helpful to express this as
\be
|{\cal M}|^2_{\rm helicity \, flip} = \frac{16 g^4 m^2}{3 s_m^4} \left[ -\frac{a}{t} + \frac{b}{u_m^2} + \frac{c}{u_m} + d  u_m - e \right]
\ee
where the coefficients
\ba
a &=& 9 s_m^4 \nonumber \\
b &=& \frac{4}{3} s_m^3 \left(s^2 + 3m^4 \right) \nonumber \\
c &=& \frac{1}{3} s_m^2 \left( 3s^2 + 25 m^2 s + 38 m^4 \right) \nonumber \\
d &=& \frac{2}{3} \left( 42 s^2 - 9 m^2 s + 2m^4 \right) \nonumber \\
e &=& \frac{1}{3} s_m \left( 7 s^2 - 133 m^2 s + 6m^4 \right) 
\ea
depend only on $s$ and $m^2$.

Integration over $t$
\be
\int_{-s_m^2/s + k_c^2}^{-k_c^2} dt \, |{\cal M}|_{{\rm helicity \, flip}}^2 = \int_{-s_m + k_c^2}^{s_m^2/s - s_m - k_c^2} du_m \, |{\cal M}|_{{\rm helicity \, flip}}^2
\ee
results in
\ba
\lefteqn{ \int_{-s_m^2/s + k_c^2}^{-k_c^2} dt \, |{\cal M}|_{{\rm helicity \, flip}}^2 = \frac{16 g^4 m^2}{3 s_m^4} \Bigg\{ a \ln\left( \frac{s_m^2}{s k_c^2} - 1 \right)
- c  \ln\left( \frac{s(s_m - k_c^2)}{s(m^2 + k_c^2) - m^4} \right) } \nonumber \\
&+& \frac{ b (s_m^2 - 2 k_c^2 s)}{(s_m-k_c^2)\left(s(m^2 + k_c^2) - m^4\right)} + \frac{d s_m}{2} \left[ \frac{s_m^3}{s^2} - \frac{2(s_m + k_c^2) s_m}{s} + 4 k_c^2 \right]
\nonumber \\ 
&-& e \left( \frac{s_m^2}{s} - 2k_c^2 \right) \Bigg\} \,.
\ea
The remaining integral over $s$ is
\ba
I_g &=& \int_{s_0}^{\infty} \frac{ds}{s}  \frac{1}{s_m^4} \Bigg\{  \frac{b(s_m^2 - 2 k_c^2 s)}{(s_m-k_c^2)\left(s(m^2 + k_c^2) - m^4\right)} + 
 \frac{d s_m}{2} \left[ \frac{s_m^3}{s^2} - \frac{2(s_m + k_c^2) s_m}{s} + 4 k_c^2 \right] \nonumber \\
&-& e \left( \frac{s_m^2}{s} - 2k_c^2 \right) + a \ln\left( \frac{s_m^2}{s k_c^2} - 1 \right) - c  \ln\left( \frac{s(s_m - k_c^2)}{s(m^2 + k_c^2) - m^4} \right)  \Bigg\}
\nonumber \\ 
&\times& \ln \left( 1 - e^{-s/4ET} \right)^{-1} \,.
\label{s-int2}
\ea
Finally we obtain the rate as
\be
\frac{1}{\tau_g(E)}  = \frac{\alpha_s^2 \, T}{3\pi} \frac{m^2}{E^2} \left( 1 + e^{-E/T} \right) I_g \,.
\ee

In principle one should use a density matrix to determine the spin or helicity relaxation time along the vorticity axis.  We shall be content to use this formula as a proxy, recognizing that it should be a very close estimate.

\section{Numerical Results}
\label{SecIX}

In this section we provide numerical results for the relaxation times using the formulas derived in previous sections.  The strange quark mass is set at $m = 110$ MeV.  We begin with vorticity fluctuations and move on to helicity flip in quark-gluon plasma.

Apart from the mass and momentum of the strange quark, the vorticity fluctutations require knowledge of the average vorticity $\omega_0$, the magnitude of the fluctutations $\overline{\omega_{\perp}^2}$, and the correlation time $\tau_c$ for these fluctuations.  These cannot be known {\it a priori} but must be found by a combination of experimental measurements and numerical simulations of high energy heavy ion collisions.  They clearly depend on the beam energy, size of the colliding nuclei, and centrality (impact parameter).  Measurements by the STAR Collaboration \cite{Nature} of the hyperon polarization indicate that $\omega_0 = (9 \pm 1) \times 10^{21}$ s$^{-1}$, with a systematic error of a factor of two, when averaging over the entire RHIC energy range.   This converts to an energy of $\omega_0 = 6$ MeV.  For illustrative purposes we take $\overline{\omega_{\perp}^2} = 8$ MeV$^2$ and $\tau_c = 4$ fm/c.  

\begin{figure}[hh]
\centering
\includegraphics[scale=0.6]{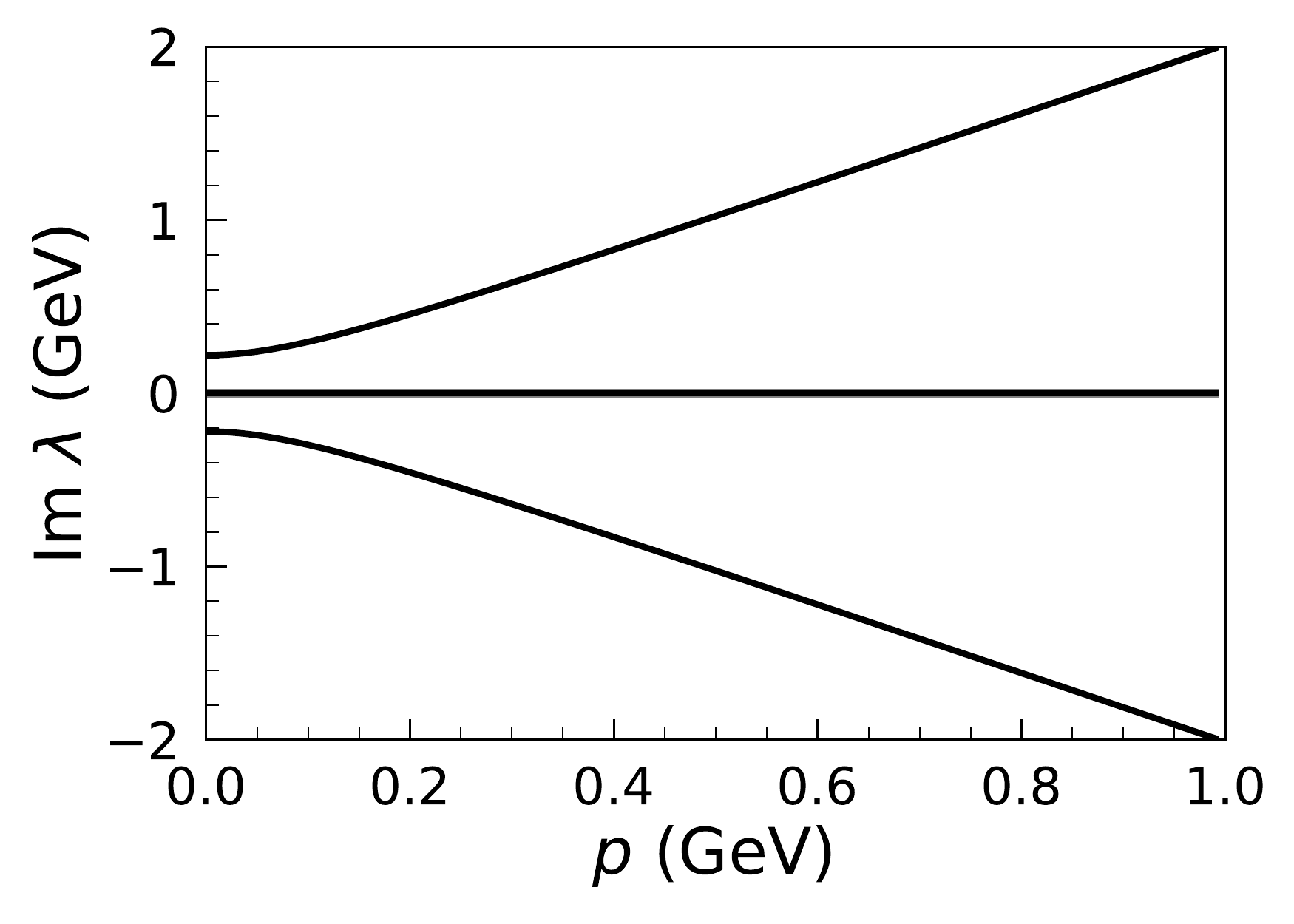}
\caption{Imaginary parts of the eigenvalues from Sec. \ref{SecVII} on vorticity fluctuations.  Two of them are zero.  The other two are closely approximated by $\pm 2 E$.}
\label{rates_im_ev_m_110-1}
\end{figure}

Figure \ref{rates_im_ev_m_110-1} shows the imaginary part of the four eigenvalues coming from vorticity fluctuations.  
Two of the eigenvalues are purely real, while the other two are complex conjugates of each other.  With the chosen parameters, and within numerical accuracy, one of the purely real eigenvalues has the eigenvector $(m/E) \langle \beta(t) \rangle + (p/E) \langle \alpha_3(t) \rangle$.  The other one has the eigenvector $\langle \Sigma_3(t) \rangle$ with an admixture of $(m/E) \langle \beta(t) \rangle + (p/E) \langle \alpha_3(t) \rangle$ at the level of $10^{-4}$ or less.  These are associated with the two operators which commute with the Hamiltonian.  The complex eigenvalues have eigenvectors which are linear combinations of $(m/E) \langle \beta(t) \rangle - (p/E) \langle \alpha_3(t) \rangle$ and 
$\langle i \beta(t) \alpha_3(t) \rangle$, whose associated operators do not commute with the Hamiltonian.  The imaginary parts are nearly equal to $\pm 2 E i$, following the discussion at the end of Sec. \ref{SecVII} in the limits $m=0$ and $p=0$. 

\begin{figure}[hh]
\centering
\includegraphics[scale=0.6]{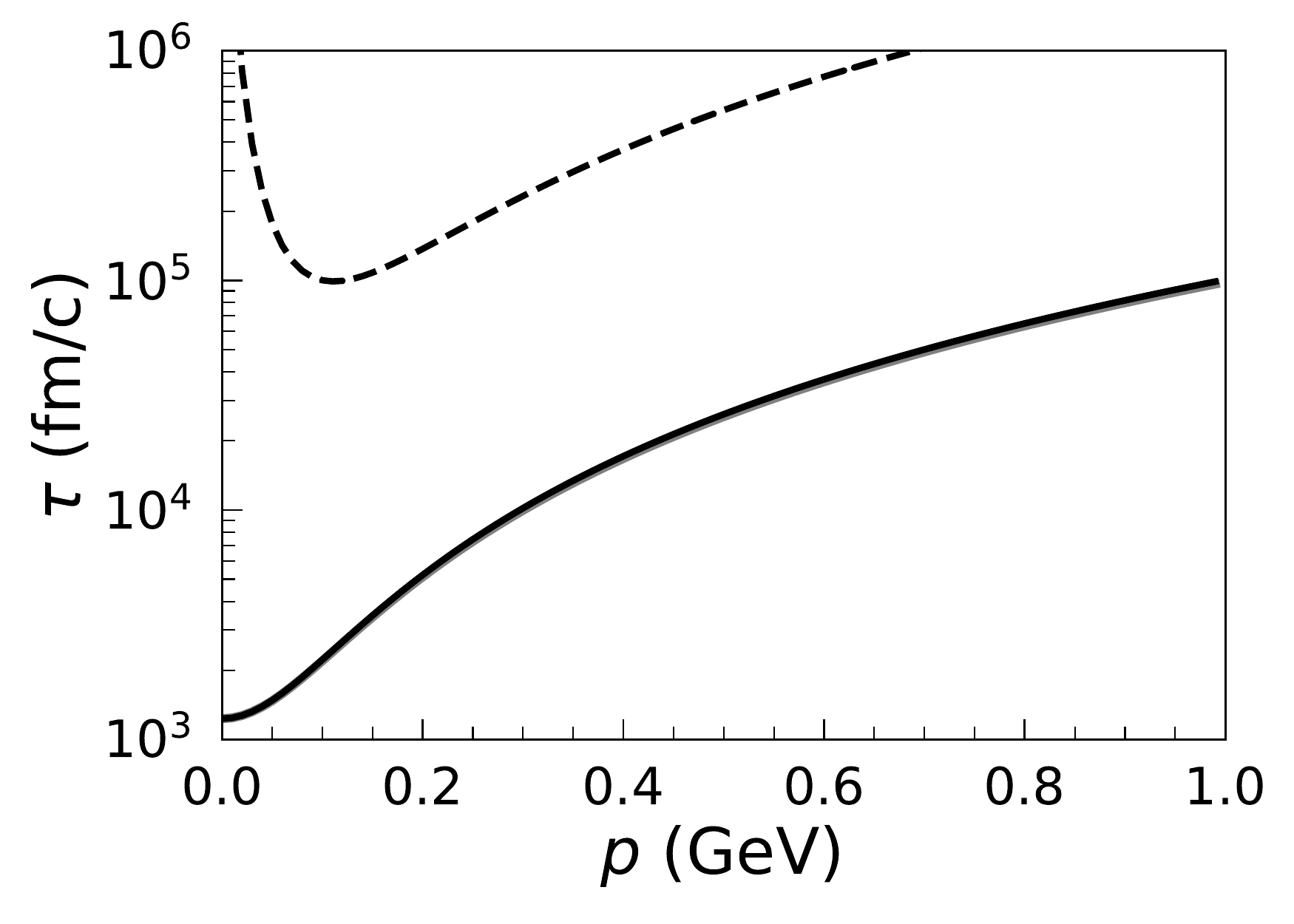}
\caption{Equilibration times for the four modes from Sec. \ref{SecVII} on vorticity fluctuations.  Three of them are nearly equal and shown as the solid curve.}
\label{rates_re_ev_m_110}
\end{figure}

Figure \ref{rates_re_ev_m_110} shows the equilibration time for the four modes.  
These come from the inverse of the real parts of the eigenvalues.  The largest equilibration time diverges like $1/p^2$ as $p \rightarrow 0$ and is associated with the zero eigenvalue of $\langle \beta(t) \rangle$, as discussed at the end of Sec. \ref{SecVII}.  This mode has a minimum when $p \approx m$, which arises from the transition from $\langle \beta(t) \rangle$ to $\langle \alpha_3(t) \rangle$ as described above.  When $p \ll m$ this eigenvalue has the limiting form
\be
\lambda \rightarrow - \frac{N}{D} \, p^2
\ee
where
\ba
N &=& 2 \left( \frac{T_- + T_+}{T_0} \right) \overline{\omega_{\perp}^2} \tau_c - 
\left( \frac{T_- + T_+}{2T_0}  -1 \right) \frac{\left(\overline{\omega_{\perp}^2} \tau_c\right)^3}{m^2} \nonumber \\
&+& \tau_c \left[ 2 \left( \frac{T_- + T_+}{T_0} \right) - \frac{\omega_0}{m} \left( \frac{T_- - T_+}{T_0} \right) \right] 
\left(\overline{\omega_{\perp}^2} \tau_c\right)^2 \,.
\ea
In this expression and this expression only the $T_{\pm}$ are evaluated at $E = m$.  Also
\be
D = 4m^2 + \left(\overline{\omega_{\perp}^2} \tau_c\right)^2 \,.
\ee
When $p \gg m$ it has the limiting form
\be
\lambda \rightarrow -  \frac{\overline{\omega_{\perp}^2} \tau_c}{1 + 4 E^2 \tau_c^2} \,.
\ee
A good representation of the dashed curve is
\be
\tau \approx \frac{D}{N} \frac{1}{p^2} + \frac{1 + 4 E^2 \tau_c^2}{\overline{\omega_{\perp}^2} \tau_c} \,.
\ee

The other three modes have smaller equilibration times and appear to all be equal, but that is only because of the logarithmic scale used in the figure.  As we saw at the end of Sec. \ref{SecVII}, these three modes become degenerate when $p \rightarrow 0$.  At that point the equilibration time is
\bd
\frac{1 + \omega_0^2 \tau_c^2}{\overline{\omega_{\perp}^2} \tau_c} \,.
\ed
Within the thickness of the solid curve in the figure they can all be approximated by the formula 
\be
\tau \approx \left( \frac{1 + 4 E^2 \tau_c^2}{\overline{\omega_{\perp}^2} \tau_c} \right) \left(  \frac{1 + \omega_0^2 \tau_c^2}{1 + 4 m^2 \tau_c^2} \right)
\ee
as a function of $p$.  Since the time scale for quark-gluon plasma expansion, cooling and entering the hadronization stage is in the range from several to at most ten fm/c, it is clear that these equilibration times are far too large to influence the evolution of strange quark spin. 

\begin{figure}[hh]
\centering
\includegraphics[scale=0.6]{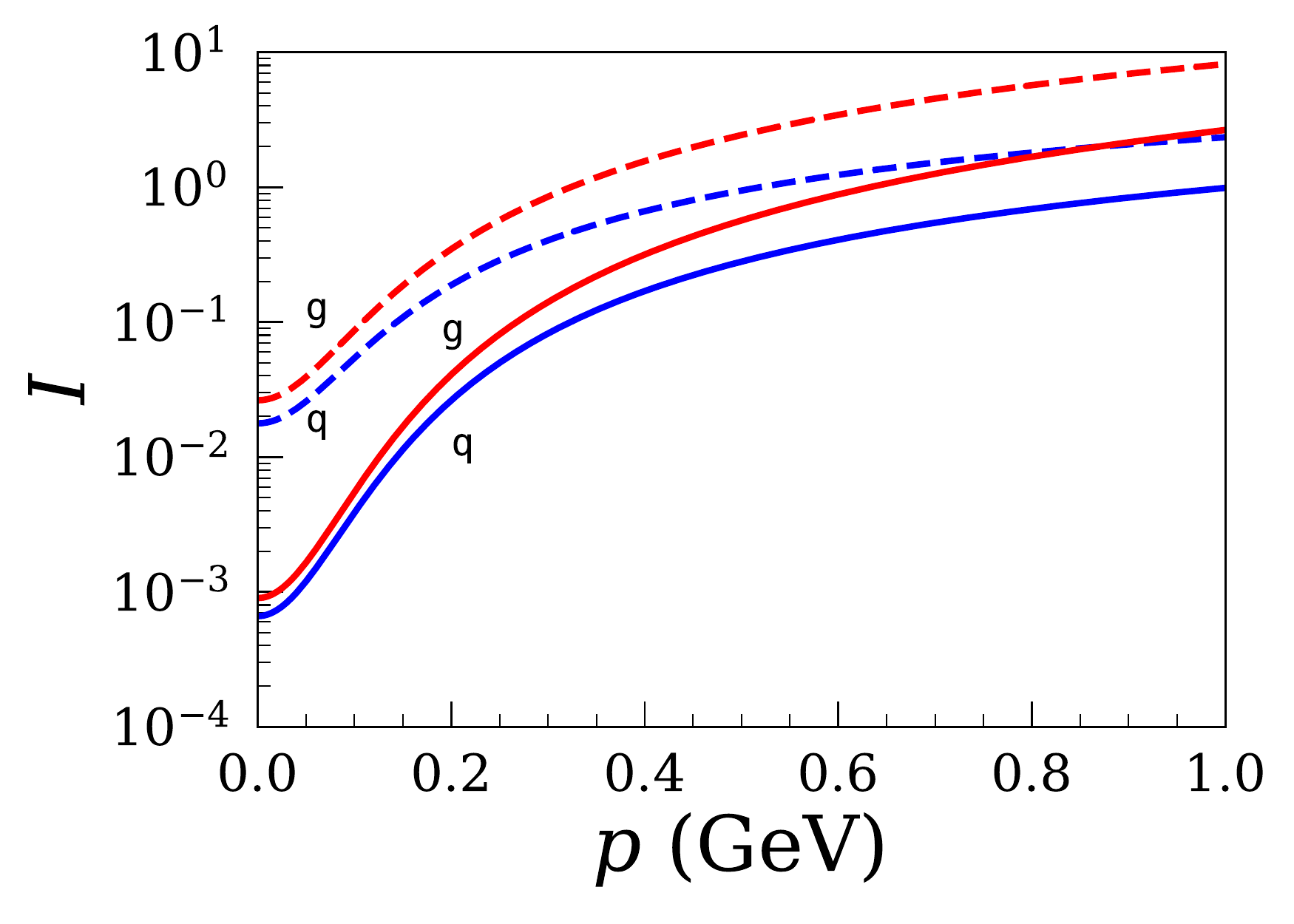}
\caption{(Color online)  The dimensionless integrals $I_q$ (blue) and $I_g$ (red) appearing in the kinetic theory.  The dashed curves correspond to a temperature of 200 MeV and the solid curves to 400 MeV.}
\label{I_q_g_m_110}
\end{figure}

For numerical estimates of the helicity flip rate we take $g=2$ corresponding to $\alpha_s = 1/\pi$.  This may not seem like a small number, and it is not, but it is a realistic number for plasma temperatures on the order of 200 and 400 MeV.  For comparison the fine structure constant 1/137 implies that the electromagnetic coupling is $e = 0.30$.  Figure \ref{I_q_g_m_110} shows the dimensionless integrals $I_q$ and $I_g$ appearing in Eqs. (\ref{s-int1}) and  (\ref{s-int2}).  Both integrals increase with momentum and decrease with temperature.  The integral coming from scattering with massless quarks is comparable to the integral coming from scattering with gluons.

\begin{figure}[hh]
\centering
\includegraphics[scale=0.6]{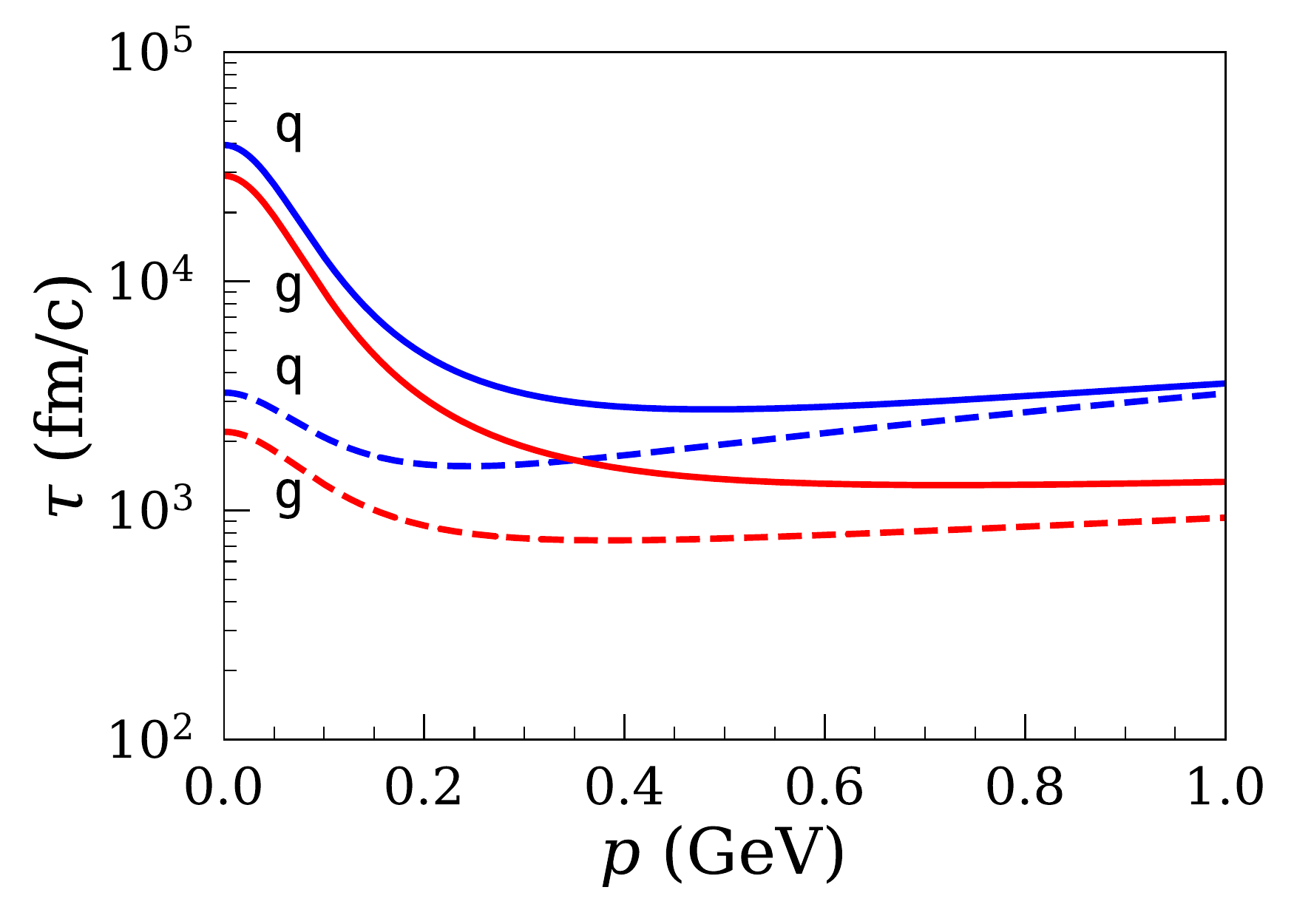}
\caption{(Color online) The equilibration times from kinetic theory for the reactions $s+q$ (blue) and $s+g$ (red).  The dashed curves correspond to a temperature of 200 MeV and the solid curves to 400 MeV.  }
\label{rates_kinetic_q_g_m_110}
\end{figure}

\begin{figure}[hh]
\centering
\includegraphics[scale=0.6]{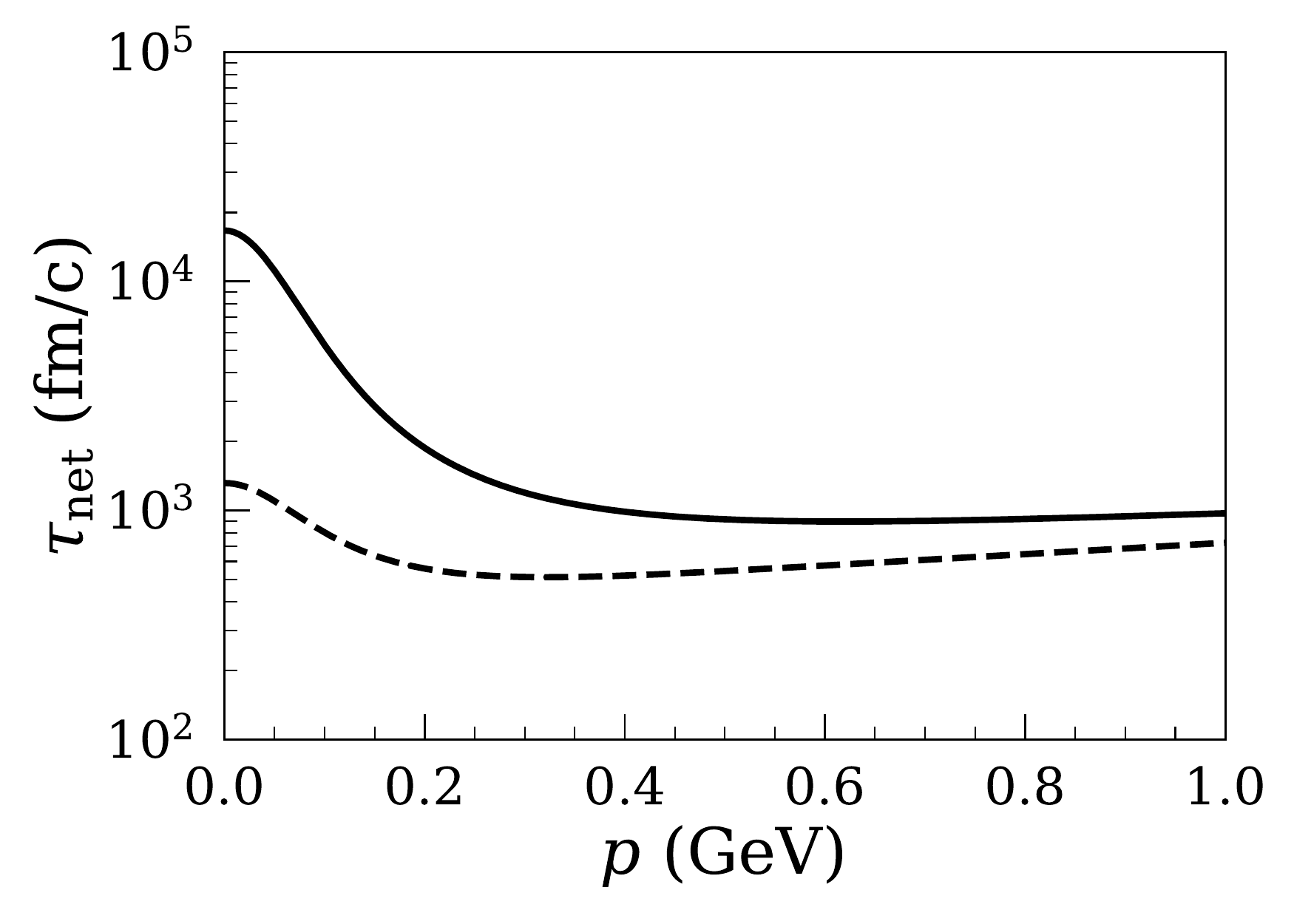}
\caption{The equilibration time from kinetic theory including both the reactions $s+q$ and $s+g$.  The dashed curve corresponds to a temperature of 200 MeV and the solid curve to 400 MeV.  }
\label{rates_kinetic_tot_m_110}
\end{figure}

Figure \ref{rates_kinetic_q_g_m_110} shows the equilibration times for strange quark helicity separately for the reactions $s+q$ and $s+g$.  At a given temperature the equilibration time for scattering with gluons is slightly smaller than scattering with massless quarks.  Figure \ref{rates_kinetic_tot_m_110} shows the net equilibration time $\tau_{\rm net}^{-1} = \tau_q^{-1} + \tau_g^{-1}$.  These equilibration times are also far too large to influence the evolution of strange quark spin in high energy heavy ion collisions. 

\newpage

\section{Conclusion}
\label{SecX}

Measurements of the net polarization of $\Lambda$ and $\bar{\Lambda}$ hyperons in heavy ion collisions at RHIC have been interpreted as being due to a coupling between their spin and the vorticity of the fluid created in these collisions.  This motivated the present study of the equilibration time for strange quark spin in rotating quark-gluon plasma, as the spin in these hyperons is generally attributed to the strange quark.  We considered two mechanisms: vorticity fluctuations and helicity flip in scatterings between strange quarks and light quarks and gluons.  Our calculations lead to equilibration times orders of magnitude too large to be relevant to heavy ion collisions.  

Certainly our calculations and parameters can be improved upon.  Regarding vorticity fluctuations, we made some rough estimates of the parameters involved, but ultimately they should come from numerical simulations of high energy heavy ion collisions.  In order to make the calculations relatively tractable for a strange quark moving with relativistic speeds we studied only the case where its momentum was either parallel or anti-parallel to the vorticity.  Otherwise we would have to include orbital angular momentum, which would probably entail using an angular momentum basis rather than a linear momentum basis.  Relaxing that restriction cannot change the equilibration time by very much, as the nonrelativistic limit yields a result independent of the direction of the momentum.  In fact, the nonrelativistic limit can be applied to the hyperons in the later hadronic phase.  Our results strongly suggest that vorticity fluctuations cannot alter the polarization of a hyperon either.

Regarding helicity flip, the contribution from hard thermal loops \cite{Kapusta-Gale} should be calculated.  In this paper we assumed that the infrared cutoff should be associated with the thermal mass of the exchanged gluon in the $t$-channel.  Given how small the helicity flip rate is makes it questionable whether such a calculation has more than theoretical interest.

How then do the $\Lambda$ and $\bar{\Lambda}$ hyperons acquire their polarization in heavy ion collisions?  One possibility is that the strange quarks are created with a net polarization.  It is generally acknowledged that gluons are produced first, and they create the majority of the quarks and anti-quarks.  Phase space alone would favor quarks and anti-quarks being created with a preference for their spin to be aligned with the vorticity.  Vorticity fluctuations and helicity flip scattering would not change the polarization as the quark-gluon plasma evolves.  The strange quarks and anti-quarks could then pass along their spin to the hyperons.  

A specific mechanism for polarizing quarks during the initial stage of heavy ion collisions was proposed in Ref. \cite{Wang1} and developed further in Refs. \cite{Wang2,Wang3}.  They considered parton-parton scattering in a longitudinal shear flow.  By Fourier transforming the differential cross section from transverse momentum to transverse distance, and taking into account the asymmetry in coordinate space due to the shear, they find that the partons can become polarized.  Recently this idea has been extended to include the scattering of wave packets \cite{Wang4}.  Whether this is in fact the origin of the polaization observed remains to be seen.  If it is, then the relationship between hyperon polarization and vorticity is much more complicated than equipartition of energy, as is usually assumed.

Recently, a general analysis of the spin current in relativistic viscous fluids has been proposed \cite{Matsuo2019}.  It points out that the spin current is not conserved, because angular momentum can be transferred between orbital and spin degrees of freedom.  As with other transport coefficients, the corresponding relaxation times are not determined by thermodynaics but must be calculated from the microscopic dynamics.

Another possibility is that strange quarks and anti-quarks in the quark-gluon plasma phase in heavy ion collisions are not polarized at all.  Rather, the hadronization from quarks and gluons to hadrons favors hyperons polarized parallel to the vorticity simply due to the available phase space.

Finally, we point out the potential application of our work to the chiral magnetic effect (CME) and the chiral vortical effect (CVE) in high energy heavy ion collisions \cite{status_report}.\\

\section*{Acknowledgement}
The work of JIK was supported by the U.S. Department of Energy Grant DE-FG02-87ER40328.  The work of ER was supported by the U.S. National Science Foundation Grant PHY-1630782 and by the Heising-Simons Foundation Grant 2017-228.


\appendix

\section{Dirac Algebra}

Following is a compendium of commutators and anticommutators useful for the matrices that frequently enter the calculations.
\ba
\left[ \alpha_i , \alpha_j \right] &= 2 i \epsilon_{ijk} \Sigma_k \;\;\;\;\;\;\;\; \left\{ \alpha_i , \alpha_j \right\} &= 2 \delta_{ij} \nonumber \\
\left[ \Sigma_i , \Sigma_j \right] &= 2 i \epsilon_{ijk} \Sigma_k \;\;\;\;\;\;\;\, \left\{ \Sigma_i , \Sigma_j \right\} &= 2 \delta_{ij} \nonumber \\
\left[ \alpha_i , \Sigma_j \right] &= 2 i \epsilon_{ijk} \alpha_k \;\;\;\;\;\;\;\,\, \left\{ \alpha_i , \Sigma_j \right\} &= 2 \delta_{ij} \gamma_5
\\
\left[ \gamma_5 , \alpha_i \right] &= 0 \;\;\;\;\;\;\;\;\;\;\;\;\;\;\;\;\;\;\; \left\{ \gamma_5 , \alpha_i \right\} &= 2 \Sigma_i \nonumber \\
\left[ \gamma_5 , \Sigma_i \right] &= 0 \;\;\;\;\;\;\;\;\;\;\;\;\;\;\;\;\;\;\; \left\{ \gamma_5 , \Sigma_i \right\} &= 2 \alpha_i \nonumber \\
\left[ \beta, \Sigma_i \right] &= 0 \,\;\;\;\;\;\;\;\;\;\;\;\;\;\;\;\;\;\;\;\; \left\{ \beta, \alpha_i \right\} &= 0 \nonumber \\
\left[ \beta \alpha_i, \alpha_j \right] &= 2 \beta \delta_{ij} \,\;\;\;\;\;\;\;\;\;\;\;\;\;\; \left\{ \beta, \gamma_5 \right\} &= 0 \nonumber \\
\left[ \beta \alpha_i, \Sigma_j \right] &= 2 i \epsilon_{ijk} \beta \alpha_k  \,\;\;\;\;\;\;\;\;\; \left[ \beta, \alpha_i \right] &= 2 \beta \alpha_i
\ea

\section{Massless Dirac Particle}

This appendix contains some of the details of the calculations in Sec. \ref{SecVI}.  Starting with the Hamiltonian
\be
H_0 = p \alpha_3 - \thalf \omega_0 \Sigma_3
\ee
it is straightforward to calculate the first few terms in the expansion of $e^{iH_0 t}$.
\ba
H_0^2 &=& \left( p^2 + \oneqt \omega_0^2\right) I - p\omega_0 \gamma_5 \nonumber \\
H_0^3 &=& p \left( p^2 + \ttqt \omega_0^2 \right) \alpha_3 - \thalf \omega_0 \left( 3p^2 + \oneqt \omega_0^2 \right) \Sigma_3 \nonumber \\
H_0^4 &=& \left( p^4 + \tthalf p^2 \omega_0^2 + {\textstyle{\frac{1}{16}}} \omega_0^4 \right) I - 2 p \omega \left( p^2 + \oneqt \omega_0^2 \right) \gamma_5
\label{H0n}
\ea
We observe that only four matrices enter the expansion.  Therefore we can express the evolution operator in the general form
\be
e^{iH_0 t} = C_1 I + C_2 \gamma_5 +i C_3 \alpha_3 + i C_4 \Sigma_3
\ee
where all $C_i$ are real.  Since the evolution operator is unitary $e^{iH_0 t} e^{-iH_0 t} = 1$ we find the two conditions
\be
C_1^2 + C_2^2  + C_3^2  +  C_4^2 = 1
\ee
and 
\be
C_1 C_2 + C_3 C_4 = 0 \,.
\ee
The first condition suggests that we express the $C_i$ in terms of the Hopf angles for a sphere in four dimensions.
\ba
C_1 &=& \cos \xi_2 \cos \eta \nonumber \\
C_2 &=& \sin \xi_1 \sin \eta \nonumber \\
C_3 &=& \cos \xi_1 \sin \eta \nonumber \\
C_4 &=& \sin \xi_2 \cos \eta \,.
\ea
The second condition says that $\xi_1 = - \xi_2 \equiv \xi$.  Hence the evolution operator can be expressed of those two angles by
\be
e^{iH_0 t} = \cos \xi \cos \eta \, I + \sin \xi \sin \eta \, \gamma_5 +i \cos \xi \sin \eta \, \alpha_3 - i \sin \xi \cos \eta \, \Sigma_3 \,.
\ee
The angles must both vanish when $t=0$.  Doing a Taylor series expansion in $\xi$ and $\eta$, and comparing to Eq. (\ref{H0n}), results in the identification $\eta = pt$ and 
$\xi = \thalf \omega_0 t$.  This is a clear generalization of Eq. (\ref{simples}).

\section{Massive Dirac Particle}

This appendix contains some of the details of the calculations in Sec. \ref{SecVII}.  Starting with the Hamiltonian
\be
H_0 = m \beta + p \alpha_3 - \thalf \omega_0 \Sigma_3
\ee
the first few terms in the expansion of $e^{iH_0 t}$ are
\ba
H_0^2 &=& E_{||}^2 I - p\omega_0 \gamma_5 - m\omega_0 \beta\Sigma_3 \nonumber \\
H_0^3 &=& m( E_{||}^2 + \thalf \omega_0^2) \beta + p \left(  E_{||}^2 + \omega_0^2 \right) \alpha_3 
- \omega_0 \left( \thalf E_{||}^2 + E^2  \right) \Sigma_3 \nonumber \\
H_0^4 &=& \left(  E_{||}^4 + \omega_0^2 E^2 \right) I - 2 p \omega_0 E_{||}^2 \gamma_5 - 2m\omega_0 E_{||}^2 \beta\Sigma_3
\ea
where $E^2 = p^2 + m^2$ and $E_{||}^2 = E^2 + \oneqt \omega_0^2$.  We observe that six matrices enter the expansion.  Hence the evolution operator has the form
\be
e^{iH_0 t} = C_1 I + C_2 \gamma_5 + C_3 \beta \Sigma_3 + iC_4 \beta + i C_5 \alpha_3 + i C_6 \Sigma_3
\ee
where all $C_i$ are real.  Requiring that the above expression be unitary results in the constraints
\ba
C_1 C_2 + C_5 C_6 &=& 0 \nonumber \\
C_1 C_3 + C_4 C_6 &=& 0 \nonumber \\
C_2 C_4 - C_3 C_5 &=& 0 \,,
\ea
and
\be
\sum_{i=1}^6 C_i^2 = 1 \,.
\ee
Matching the Taylor series expansion of the exponential results in Eq. (\ref{massiveCs}).

Operators in the interaction picture can be obtained from those in the Schr\"odinger picture.  We shall need
\be
\Sigma_1(t) = B_1(t) \Sigma_1 + B_2(t) \Sigma_2
+ B_3(t) \alpha_1 - B_4(t) \alpha_2 + i B_5(t) \beta \alpha_1 + i B_6(t) \beta \alpha_2 \,,
\ee
\be
\Sigma_2(t) = B_1(t) \Sigma_2 - B_2(t) \Sigma_1
+ B_3(t) \alpha_2 + B_4(t) \alpha_1 + i B_5(t) \beta \alpha_2 - i B_6(t) \beta \alpha_1 \,,
\ee 
while $\Sigma_3(t) = \Sigma_3$ on account of the fact that it commutes with $H_0$.  The $B_i$ are expressed in terms of the $C_i$ as
\ba
B_1 &=&  C_1^2 + C_2^2 - C_3^2 + C_4^2 - C_5^2 - C_6^2 \nonumber \\
B_2 &=&  2 \left( C_3 C_4 - C_1 C_6 - C_2 C_5 \right) \nonumber \\
B_3 &=&  4 C_1 C_2 \nonumber \\
B_4 &=&  2 \left( C_1 C_5 + C_2 C_6 \right) \nonumber \\
B_5 &=&  4 C_2 C_4 \nonumber \\
B_6 &=&  2 \left( C_2 C_3 - C_4 C_5 \right) \,.
\ea
When averaging over fluctuations it is helpful to express these as sums of sines and cosines.
\ba
B_1 &=& \frac{p^2}{2E^2} \left[ \cos\left( (2E+\omega_0)t \right) + \cos\left( (2E-\omega_0)t \right) \right] + \frac{m^2}{E^2} \cos(\omega_0 t) \nonumber \\
B_2 &=& \frac{p^2}{2E^2} \left[ \sin\left( (2E+\omega_0)t \right) - \sin\left( (2E-\omega_0)t \right) \right] + \frac{m^2}{E^2} \sin(\omega_0 t) \nonumber \\
B_3 &=& \frac{p}{2E} \left[ \cos\left( (2E-\omega_0)t \right) - \cos\left( (2E+\omega_0)t \right) \right] \nonumber \\
B_4 &=& \frac{p}{2E} \left[ \sin\left( (2E-\omega_0)t \right) + \sin\left( (2E+\omega_0)t \right) \right] \nonumber \\
B_5 &=& \frac{pm}{2E^2} \left[ \sin\left( (2E-\omega_0)t \right) - \sin\left( (2E+\omega_0)t \right) + 2 \sin(\omega_0 t)\right]  \nonumber \\
B_6 &=& \frac{pm}{2E^2} \left[ \cos\left( (2E+\omega_0)t \right) + \cos\left( (2E-\omega_0)t \right) - 2\cos(\omega_0 t) \right] 
\ea

Similarly one finds that
\ba
\alpha_3(t) &=& \left( C_1^2 + C_2^2 - C_3^2 - C_4^2 + C_5^2 + C_6^2 \right) \alpha_3 + 2 \left( C_2 C_3 + C_4 C_5 \right) \beta 
+  2 i (C_1 C_4 - C_3 C_6) \beta \alpha_3 \nonumber\\
&=&  \left[ \frac{p^2}{E^2} + \frac{m^2}{E^2}  \cos(2Et) \right] \alpha_3 + \frac{mp}{E^2} \left[ 1 - \cos(2Et) \right] \beta
+ i \frac{m}{E} \sin(2Et) \beta \alpha_3 \,,
\ea
and
\ba
\beta(t) &=& \left( C_1^2 - C_2^2 + C_3^2 + C_4^2 - C_5^2 + C_6^2 \right) \beta + 2 \left( C_2 C_3 + C_4 C_5 \right) \alpha_3 
+  2 i (C_2 C_6 - C_1 C_5) \beta \alpha_3 \nonumber \\
&=& \left[ \frac{m^2}{E^2} + \frac{p^2}{E^2} \cos(2Et) \right] \beta + \frac{mp}{E^2} \left[ 1 - \cos(2Et) \right] \alpha_3
- i \frac{p}{E} \sin(2Et) \beta \alpha_3 \,.
\ea
The inverse relations are
\ba
  \beta&=& \left( \frac{m^2}{E^2} +\frac{p^2}{E^2} \cos (2Et ) \right) \beta(t)
  + \frac{m p}{E^2} \left( 1 - \cos(2Et)\right)  \alpha_3(t) + i  \frac{p}{E} \sin (2 E t) \beta(t) \alpha_3(t) \nonumber \\
  \alpha_3 &=& \frac{m p}{E^2} \left( 1 - \cos(2Et)\right) \beta(t) + \left( \frac{p^2}{E^2} + \frac{m^2}{E^2} \cos (2 E t ) \right) \alpha_3(t) - i \frac{m}{E} \sin (2 E t) \beta(t) \alpha_3(t) \nonumber \\
  \beta \alpha_3&=&  i \frac{p}{E} \sin (2 E t) \beta(t) -i \frac{m}{E} \sin (2 E t) \alpha_3(t) + \cos(2 E t) \beta(t) \alpha_3(t) \nonumber \\
  \Sigma_3 &=& \Sigma_3(t) \,.
\ea

\section{Quark Helicity Flip Amplitude}

Consider a massless quark scattering from a massive strange quark.  To lowest order there is only a single Feynman diagram which involves the exchange of a gluon in the $t$ channel.  We are interested in the situation where the strange quark changes its helicity.  The helicity of the massless quark cannot change.  The amplitude is denoted by 
${\cal M}(\sigma \sigma', -\sigma \sigma')$ where $\sigma$ is the helicity of the incoming strange quark and $\sigma'$ is the helicity of the massless quark.  Using the method of Ref. \cite{Fearing} one readily finds that
\be
{\cal M}(\sigma \sigma', -\sigma \sigma') = \frac{2 g^2 m p_* \sin \theta_*}{t} T^a_{ji} T^a_{lk}
\ee
irrespective of the sign of $\sigma$ and $\sigma'$.  Here $p_*$ is the momentum and $\theta_*$ the scattering angle in the center of momentum frame, and $i, j, k, l$ are the quark colors.  Squaring this, averaging over initial colors and summing over all colors using
\be
\sum_{a,b} \left( {\rm Tr} \, T^a T^b \right)^2 = \oneqt (N^2 - 1)
\ee
for an $SU(N)$ gauge theory, gives
\be
|{\cal M}|_{{\rm helicity \, flip}}^2 = \frac{8}{9} g^4 \frac{m^2 p_*^2}{t^2} \sin^2\theta_* = - \frac{8}{9} g^4 m^2 \left[ \frac{1}{t} + \frac{s}{(s-m^2)^2} \right]
\ee
for any choice of $\sigma$ and $\sigma'$.  This result is also true for scattering of a massless anti-quark.

To our knowledge the quark helicity flip amplitude for a massive quark scattering with a gluon has not been published before.  Here we outline the major steps of the calculation at the tree level. The amplitude for scattering of a quark of helicity $\sigma$ with a gluon of helicity $\lambda$ to a quark of helicity $\sigma'$ and a gluon of helicity $\lambda'$ is denoted by
${\cal M}(\sigma \lambda, \sigma' \lambda')$.  It is useful to define $s_m = s - m^2$, $u_m = u - m^2$, and $r^2 = m^4 - s u$.  Then $s_m + t + u_m = 0$.  We use the polarization vector choice of Ref. \cite{DD}.  After a lengthy calculation one finds the amplitudes
\ba
{\cal M}(++,-+) &=& {\cal M}(+-,--) = \frac{2 g^2 m r^2 \sqrt{-t}}{s_m^2}
\left[ \frac{1}{2u_m} \{T^a,T^b\}_{ji} + \left( \frac{1}{2u_m} - \frac{1}{t} \right) i f^{abc} T^c_{ji} \right] \nonumber \\
{\cal M}(++,--) &=& \frac{g^2 m \sqrt{-t}}{s_m^2} 
\left[ \frac{m^2 t}{u_m} \{T^a,T^b\}_{ji} + \left( \frac{r^2 - s_m u_m}{u_m} - \frac{2r^2}{t} \right) i f^{abc} T^c_{ji} \right] \nonumber \\
{\cal M}(+-,-+) &=& \frac{2 g^2 st \sqrt{-t}}{s_m^2}
\left[ \frac{1}{2u_m} \{T^a,T^b\}_{ji} + \left( \frac{1}{2u_m} + \frac{1}{t} \right) i f^{abc} T^c_{ji} \right] \,.
\ea
As usual, $a$ and $b$ are the color indices for the initial and final state gluons, while $i$ and $j$ are the color indices for the initial and final state strange quark.  The only $t$-channel contribution comes from the diagram with a triple gluon vertex.  The Abelian version of these amplitudes correspond to helicity flip of an electron in Compton scattering \cite{DD}.

The squared amplitudes are
\ba
|{\cal M}(++,-+)|^2 &=& |{\cal M}(+-,--)|^2 = \frac{4 g^4 m^2 r^4 (-t)}{s_m^4} \left[ \frac{7}{3u_m^2} + 3 \left( \frac{1}{u_m} - \frac{2}{t} \right)^2 \right]
\nonumber \\
|{\cal M}(++,--)|^2 &=& \frac{4g^4 m^2 (-t)}{s_m^4} \left[ \frac{7 m^4 t^2}{3 u_m^2} + 3 \left( \frac{r^2 - s_m u_m}{u_m} - \frac{2r^2}{t} \right)^2 \right]
\nonumber \\
|{\cal M}(+-,-+)|^2 &=& \frac{4 g^4 m^2 s^2 (-t)^3}{s_m^4} \left[ \frac{7}{3u_m^2} + 3 \left( \frac{1}{u_m} + \frac{2}{t} \right)^2 \right] \,.
\ea
These are obtained by summing over all colors in the initial and final states using the trace identities
\ba
\sum_{a,b} {\rm Tr} \, \{T^a, T^b\} \{T^a, T^b\} &=& \frac{(N^2 - 1) (N^2 - 2)}{2N} = \frac{28}{3} \nonumber \\
\sum_{a,b,c,d} {\rm Tr} \, f^{abc} f^{abd} T^c T^d &=& \frac{N(N^2 - 1)}{2} = 12
\ea
where the numbers on the far right side are for $N=3$.  To obtain the rate for scattering of a strange quark with a specified (average) color we must divide these by 3.  These add incoherently, so that after addition and division by 3 we obtain
\ba
|{\cal M}|^2_{\rm helicity \, flip} &=& \frac{4 g^4 m^2 (-t)}{3 s_m^4} \Bigg\{ 4 s^2 \left[ 3 + \frac{3 t}{u_m} + \frac{4 t^2}{3 u_m^3} \right] + 
8 (m^4 - s u)^2 \left[ \frac{3}{t^2} - \frac{3}{u_m t} + \frac{4}{3 u_m^2} \right] \nonumber \\
&+& \frac{7 m^4 t^2}{3 u_m^2} + 3 \left[ \frac{(m^4 - s u - s_m u_m)}{u_m} - \frac{2(m^4 - s u)}{t} \right]^2 \Bigg\} \,.
\ea

\end{document}